\documentclass[a4paper, onecolumn, accepted=2022-12-04]{quantumarticle}

\pdfoutput=1

\usepackage[utf8]{inputenc}
\usepackage[final]{hyperref}
\usepackage[stable]{footmisc}
\usepackage{todonotes}
\hypersetup{breaklinks=true, colorlinks = true, allcolors = blue}
\usepackage{amsmath,amssymb,amsfonts, mathtools, braket}
\usepackage{amsthm, thmtools, thm-restate}
\usepackage{environ,fullpage,bm}
\usepackage{authblk}
\usepackage{caption}
\usepackage{enumitem}

\newtheorem{theorem}{Theorem} 
\newtheorem{proposition}[theorem]{Proposition}
\newtheorem{lemma}[theorem]{Lemma}

\newtheorem{corollary}[theorem]{Corollary}
\newtheorem{definition}{Definition}
\newtheorem*{remark*}{Remark(s)}

\newcommand{\Tr}{\mbox{\rm Tr}}
\newcommand{\tr}[1]{\Tr\big[#1\big]}
\newcommand{\sign}[1]{\mathrm{sign}\big(#1\big)}
\newcommand{\bigO}[1]{\mathcal{O}\big( #1 \big)}
\newcommand{\inp}[2]{\langle #1, #2\rangle}
\newcommand{\C}{\mathbb{C}}
\newcommand{\R}{\mathbb{R}}
\newcommand{\fat}{\mathrm{fat}}
\renewcommand{\epsilon}{\varepsilon}

\makeatletter
\AtBeginDocument{\let\LS@rot\@undefined}
\makeatother

\begin{document}

\title{Structural risk minimization for quantum linear classifiers}

\author{Casper Gyurik}
\email[]{c.f.s.gyurik@liacs.leidenuniv.nl}
\affiliation{LIACS, Leiden University, Niels Bohrweg 1, 2333 CA Leiden, Netherlands}

\author{Dyon van Vreumingen}
\affiliation{LIACS, Leiden University, Niels Bohrweg 1, 2333 CA Leiden, Netherlands}
\affiliation{QuSoft, Centrum Wiskunde \& Informatica (CWI), Science Park 123, 1098 XG Amsterdam, Netherlands}
\affiliation{Institute of Physics, University of Amsterdam, Science Park 904, 1098 XH Amsterdam, the Netherlands}

\author{Vedran Dunjko}
\affiliation{LIACS, Leiden University, Niels Bohrweg 1, 2333 CA Leiden, Netherlands}
\affiliation{LION, Leiden University, Niels Bohrweg 2, 2333 CA Leiden, Netherlands}

\date{December 4th, 2022}

\begin{abstract}
  Quantum machine learning (QML) models based on parameterized quantum circuits are often highlighted as candidates for quantum computing's near-term ``killer application''. 
  However, the understanding of the empirical and generalization performance of these models is still in its infancy. 
  In this paper we study how to balance between training accuracy and generalization performance (also called structural risk minimization) for two prominent QML models introduced by Havl{\'\i}{\v{c}}ek~et~al.~\cite{havlivcek:qsvm}, and Schuld and Killoran~\cite{schuld:qsvm}. 
  Firstly, using relationships to well understood classical models, we prove that two model parameters -- i.e., the dimension of the sum of the images and the Frobenius norm of the observables used by the model -- closely control the models’ complexity and therefore its generalization performance. 
  Secondly, using ideas inspired by process tomography, we prove that these model parameters also closely control the models' ability to capture correlations in sets of training examples. 
  In summary, our results give rise to new options for structural risk minimization for QML models.
\end{abstract}

\maketitle

\section{Introduction}
\label{sec:introduction}

After years of efforts the first proof-of-principle quantum computations that arguably surpass what is feasible with classical supercomputers have been realized~\cite{google:supremacy}.
As the leap from noisy intermediate-scale quantum (NISQ) devices~\cite{preskill:nisq} to full-blown quantum computers may require further decades, finding practically useful NISQ-suitable algorithms is becoming increasingly important.
It has been argued that some of the most promising NISQ-suitable algorithms are those that rely on \emph{parameterized quantum circuits} (also called variational quantum circuits)~\cite{cerezo:vqa, mcclean:vqa}.
Such algorithms have been proposed for quantum chemistry~\cite{kandala:vqe, omalley:vqe}, for optimization~\cite{farhi:qaoa}, and for machine learning~\cite{benedetti:pqcs}.
One of the advantages of parameterized quantum circuits is that restrictions of NISQ devices can be hardwired into the circuit.
Moreover, families of parameterized quantum circuits can -- under widely believed complexity-theoretic assumptions -- realize input-output correlations that are intractable for classical computation~\cite{terhal:advantage_svm, bremner:advantage}.
In this paper we discuss the application of parameterized quantum circuits as machine learning models in hybrid quantum-classical methods.
The use of NISQ devices in the context of machine learning is particularly appealing as machine learning algorithms may be more tolerant to noise in the quantum hardware~\cite{grant:noise, riste:noise}. 
In short, parameterized quantum circuits could yield NISQ-friendly machine learning models that could be used to classify data for which conventional classical machine learning models may struggle.

\begin{figure*}[h!]
\centering
\includegraphics[width=0.8\linewidth]{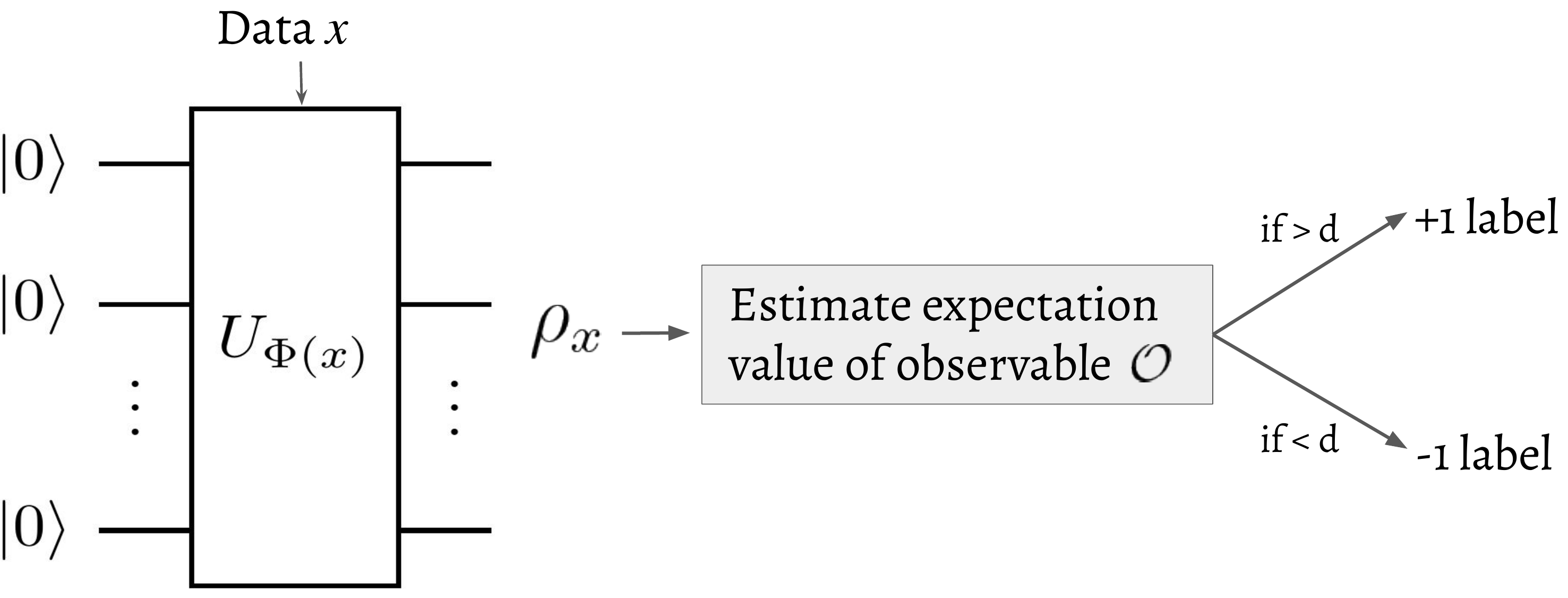}%
\captionsetup{justification=centering}
\caption{An overview of the quantum machine learning models introduced in~\cite{havlivcek:qsvm, schuld:qsvm}. 
First, a parameterized quantum circuit is used to encode the data into a quantum state $\rho_x$. 
Afterwards, an observable $\mathcal{O}$ is measured.
If its expectation value lies above $d$, then we assign the label $+1$, and $-1$ otherwise.
The goal in training is the find the optimal observable $\mathcal{O}$ and threshold $d$.
\label{fig:intro_overview}}
\end{figure*}

In machine learning, parameterized quantum circuits can serve as a parameterized family of real-valued functions in a manner similar to neural networks (they are often called quantum neural networks).
It has been noted that machine learning models based on parameterized quantum circuit are closely related to \emph{linear classifiers}, which use hyperplanes to separate classes of data embedded in a vector space.
This connection was first established by Havlí\v{c}ek et al.~\cite{havlivcek:qsvm}, and Schuld \& Killoran~\cite{schuld:qsvm}, who both defined two machine learning models based on parameterized quantum circuits that efficiently implement certain families of linear classifiers -- an illustration of which can be found in Figure~\ref{fig:intro_overview}.
In this paper we further investigate and exploit this relation between machine learning models based on parameterized quantum circuits and standard linear classifiers to investigate how to perform \emph{structural risk minimization}.
More specifically, we study how to tune parameters of quantum machine learning models to optimize their expressivity (i.e., the ability to correctly capture correlations in sets of training examples) while preventing the model from becoming too complex (which could cause it to overfit and generalize poorly to unseen examples). 

\paragraph{Our contributions}

We identify model parameters that can be used to implement structural risk minimization for a family of quantum machine learning models that includes the models introduced by Havlí\v{c}ek et al.~\cite{havlivcek:qsvm}, and Schuld \& Killoran~\cite{schuld:qsvm}.
Specifically, we theoretically analyze the effect that limiting the rank or Frobenius norm of the observables measured by these models has on the trade-off between 1.\ generalization performance, and 2.\ expressivity, and we show that:
\begin{enumerate}[label=\arabic*.]
    \item 
    \begin{enumerate}[label=(\alph*)]
        \item A measure of complexity called the \emph{VC dimension} can be controlled by limiting the \emph{dimension of the sum of the images} of the observables measured by the quantum model. 
        In particular, we provide explicit analytical bounds on the VC dimension in terms of the dimension of the span of the observables, and the dimension of the sum of the images of the observables.
        Afterwards, we use this result to devise quantum models for which we can control this VC dimension bound by limiting the ranks of the observables (i.e., they can be regularized by penalizing high-rank observables).
        \item A measure of complexity called the \emph{fat-shattering dimension} can be controlled by limiting the \emph{Frobenius norm} of the observables measured by the quantum model.
        In particular, we provide explicit analytical bounds on the fat-shattering dimension in terms of the Frobenius norm of the observable.
    \end{enumerate}
    
    Due to the well-established connection between these complexity measures and upper bounds on the generalization performance~\cite{mohri:foundations, bartlett:generalization1}, our results theoretically quantify the effect that adjusting the dimension of the sum of the images, or Frobenius norm of the observables measured by the quantum model have on its generalization performance. Further we show~that:
    \item 
    \begin{enumerate}[label=(\alph*)]
        \item Quantum models that use high-rank observables are strictly more expressive than quantum models that use low-rank observables.
        In particular, we show that i) any set of examples that can be correctly classified using a low-rank observable can also be correctly classified using a high-rank observable, and ii) there exist sets of examples that can only be correctly classified using an observable of at least a certain rank.
        \item Quantum models that use observables with large Frobenius norms can achieve larger margins (i.e., empirical quantities measured on a set of training examples that influence certain generalization bounds) compared to quantum models that use observables with small Frobenius norms.
        In particular, we show that there exist sets of examples that can only be classified with a given margin using observables of at least a certain Frobenius norm.
        Since the Frobenius norm controls the fat-shattering dimension, this can actually also have a positive effect on the generalization performance (as discussed in Section~\ref{subsec:srm}).
    \end{enumerate}
    
    To summarize, we show that the rank or Frobenius norm of the observables measured by the quantum model also controls the model's ability to capture correlations in sets of examples.
\end{enumerate}

Additional to the above two points, we also connect quantum machine learning with parameterized quantum circuits to standard structural risk minimization theory and discuss how to use our results to find the best quantum models in practice.
In particular, we discuss different types of regularization that are theoretically motivated by our results, which help improve the performance of the quantum models in practice without putting extra requirements on the quantum hardware and are thus NISQ-suitable.
Moreover, we find that there exist training methods -- i.e., those who penalize high-rank observables -- that are theoretically motivated by our results, and for which the resulting quantum model does not necessarily correspond to a kernel method as argued in~\cite{schuld:kernel}.

\paragraph{Related work} 
The way the observable in Figure~\ref{fig:intro_overview} is measured typically consists of multiple steps that involve different parts of the quantum model.
For instance, a prominent approach consists of first applying a parameterized quantum circuit to the data encoding state $\rho_x$, and then performing some fixed measurement.
Previous works have focused on showing how complexity measures depend on the different parts of the quantum model that implement the observable measurement, such as the quantum circuit ansatz~\cite{caro:power, bu:power1, bu:power2}, or the level of noise in the model~\cite{bu:power3}.
In this work we study the observable measured by the quantum model as a whole due to the 1-1 correspondence with the normal vectors of separating hyperplanes of linear classifiers.
By studying the observable as a whole, our results apply to all quantum models that are of the structure described in Figure~\ref{fig:intro_overview}, independent of how the observable measurement is implemented.
Moreover, by being agnostic to how the observable is measured, our results are complementary to results that focus on the specifics of a particular implementation of the observable measurement such as those mentioned above.
Other related work has focused on showing that quantum machine learning models are remarkably expressive and satisfy generalization bounds based on different complexity measures~\cite{abbas:power, du:power, banchi:power}. 
Finally, other related work has studied the generalization performance of quantum machine learning models in order to compare their performance with classical machine learning models~\cite{huang:power1, liu:power_appendix}.

\paragraph{Organization}
In Section~\ref{sec:background}, we define the quantum machine learning models studied in this paper, and we provide background on structural risk minimization.
In Section~\ref{sec:srm_for_qlc}, we investigate how structural risk minimization can be achieved for the quantum models. 
First, we determine two capacity measures of the quantum models, which allows us to identify model parameters that control the model's complexity in Subsection~\ref{subsec:fs-dim}.
Afterwards, we investigate the effect of these model parameters on the empirical performance in Subsection~\ref{subsec:expressivity}.
We end with a discussion of how to implement structural risk minimization of the quantum models in practice in Subsection~\ref{subsec:srm_practice}.

\section{Background and motivation}
\label{sec:background}

In this section we provide the necessary background and we motivate our results.
First, we introduce the family quantum machine learning models that we will study.
Afterwards, we introduce the framework of statistical learning theory, which together with our results will provide an approach to optimally tuning the family of quantum models via so-called \emph{structural risk minimization}.

\subsection{Quantum linear classifiers}
\label{sec:qlc}

A fundamental family of classifiers used throughout machine learning are those constructed from \emph{linear functions}.
Specifically, these classifiers are constructed from the family of real-valued functions on $\R^\ell$ given by
\begin{align}
    \label{eq:f_lin}
    \mathcal{F}_{\text{lin}} = \Big\{f_{w}(x) = \inp{w}{x} \text{ }\Big|\text{ } w \in \R^\ell  \Big\},
\end{align}
where $\inp{.}{.}$ denotes an inner product on the input space $\R^\ell$.
These linear functions are turned into classifiers by adding an offset and taking the sign, i.e., the classifiers are given by 
\begin{align}
    \label{eq:c_lin}
    \mathcal{C}_{\text{lin}} = \Big\{ c_{w, d}(x)=\sign{\inp{w}{x}-d} \text{ }\Big|\text{ } w \in \R^\ell,\text{ }d \in \R \Big\}.
\end{align}
These linear classifiers essentially use hyperplanes to separate the different classes in the data.

While this family of classifiers seems relatively limited, it becomes powerful when introducing a \emph{feature map}.
Specifically, a feature map $\Phi: \R^\ell \rightarrow \R^N$ is used to (non-linearly) map the data to a (much) higher-dimensional space -- called the \emph{feature space} -- in order to make the data more linearly-separable. 
We let $\mathcal{C}\left(\Phi\right) = \{c \circ \Phi \mid c \in \mathcal{C}\}$ denote the family of classifiers on $\R^\ell$ obtained by combining a family of linear classifiers $\mathcal{C} \subseteq \mathcal{C}_{\mathrm{lin}}$ on $\R^N$ with a feature map $\Phi$.
If the feature map is clear from the context we will omit it in the notation and just write $\mathcal{C}$. 
A well known example of a model based on linear classifiers is the \emph{support vector machine} (SVM), which aims to finds the hyperplane that attains the maximal perpendicular distance to the two classes of points in the two distinct half-spaces (assuming the feature map makes the data linearly separable).

The linear-algebraic nature of linear classifiers makes them particularly well-suited for quantum treatment.
In the influential works of Havlí\v{c}ek et al.~\cite{havlivcek:qsvm}, and Schuld \& Killoran~\cite{schuld:qsvm}, the authors propose a model where the space of $n$-qubit Hermitian operators -- denoted $\mathrm{Herm}\left(\C^{2^n}\right)$ -- takes the role of the feature space.
Specifically, they view $\mathrm{Herm}\left(\C^{2^n}\right)$ as a $4^n$-dimensional real vector space equipped with the Frobenius inner product $\inp{A}{B} = \Tr[A^\dagger B]$.
Their feature map maps classical inputs $x$ to $n$-qubit density matrices $\Phi(x) = \rho_\Phi(x)$ (i.e., positive semi-definite matrices of trace one).
Finally, the hyperplanes that separates the states $\rho_{\Phi(x)}$ corresponding to the different classes are given by $n$-qubit observables.
In short, the family of functions their model uses is given by

\begin{align}
\label{eq:f_qlin}
    \mathcal{F}_{\text{qlin}} =  \Big\{f_{\mathcal{O}}(x) = \tr{\mathcal{O}\rho_\Phi(x)} \text{ }\Big|\text{ } \mathcal{O} \in \mathrm{Herm}\big(\C^{2^n}\big)\Big\},
\end{align}
and the family of classifiers -- which we refer to as \emph{quantum linear classifiers} -- is given by
\begin{align}
\label{eq:c_qlim}
    \mathcal{C}_{\text{qlin}}= \Big\{c_{\mathcal{O}, d}(x)=\sign{\tr{\mathcal{O}\rho_\Phi(x)}-d} \text{ }\Big|\text{ }\mathcal{O} \in \mathrm{Herm}\big(\C^{2^n}\big),\text{ } d \in \R \Big\}.
\end{align}
We can estimate $f_{\mathcal{O}}(x)$ defined in Equation~\eqref{eq:f_qlin} by preparing the state $\rho_{\Phi(x)}$ and measuring the observable $\mathcal{O}$.
In particular, approximating $f_{\mathcal{O}}(x)$ up to additive error $\epsilon$ requires only $\bigO{1/\epsilon^2}$ samples.
While the error creates a fuzzy region around the decision boundary, this turns out to not cause major problems in practical settings~\cite{benedetti:pqcs}.

\begin{figure*}[h!]
\includegraphics[width=\linewidth]{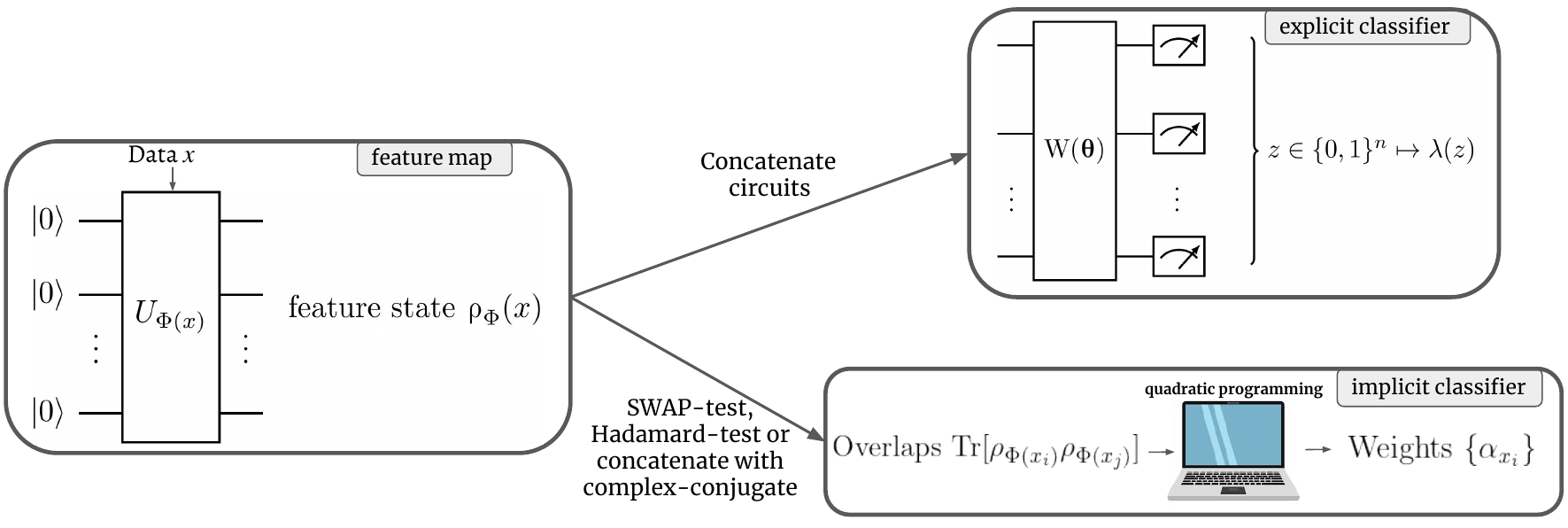}%
\caption{\label{fig:overview} An overview of the implementations of the explicit and implicit quantum linear classifiers defined in Equations~\eqref{eq:c_explicit} and~\eqref{eq:c_implicit}, respectively. 
Note that in the case of the explicit classifier, a universal circuit $W(\theta)$ (specifying the eigenbasis) followed by a computational basis measurement and universal postprocessing $\lambda$ (specifying the eigenvalues) allows one to measure any observable.}
\end{figure*}

Using parameterized quantum circuits both the preparation of a quantum state that encodes the classical input and the measurement of observables can be done efficiently for certain feature maps and families of observables.
We now briefly recap two ways in which parameterized quantum circuits can be used to efficiently implement a family of quantum linear classifiers, as originally proposed by Havlí\v{c}ek et al.~\cite{havlivcek:qsvm}, and Schuld \& Killoran~\cite{schuld:qsvm}. 
Both ways use a parameterized quantum circuit to implement the feature map.
Specifically, let $U_\Phi$ be a parameterized quantum circuit, then we can use it to implement the feature map given by
\begin{align}
\label{eq:featuremap}
\Phi: x \mapsto \rho_{\Phi}(x) \coloneqq \ket{\Phi(x)}\bra{\Phi(x)},
\end{align}
where $\ket{\Phi(x)} \coloneqq U_\Phi(x)\ket{0}^{\otimes{n}}$.
The key difference between the two approaches is which observables they are able to implement (i.e., which separating hyperplanes they can represent) and how the observables are actually measured (i.e., how the functions $f_{\mathcal{O}}$ are evaluated).
An overview of how the two approaches implement quantum linear classifiers can be found in Figure~\ref{fig:overview}, and we discuss the main ideas behind the two approaches below.

\paragraph{Explicit quantum linear classifier\footnote{Also called the \emph{quantum variational classifier}~\cite{havlivcek:qsvm}.}}
The observables measured in this approach are implemented by first applying a parameterized quantum circuit $W(\theta)$, followed by a computational basis measurement and postprocessing of the measurement outcome $\lambda : [2^n] \rightarrow \R$. 
Upon closer investigation, one can derive that the corresponding observable is given by
\begin{align}
\label{eq:otheta}
\mathcal{O}^\lambda_\theta = W^\dagger(\theta)\cdot \mathrm{diag}\Big(\lambda(0), \lambda(1), \dots, \lambda(2^n - 1)\Big)\cdot W(\theta).
\end{align}
Examples of efficiently computable postprocessing functions $\lambda$ include functions with a polynomially small support (implemented using a lookup table), functions that are efficiently computable from the input bitstring (e.g., the parity of the bitstring, which is equivalent to measuring $Z^{\otimes n}$), or parameterized functions such as neural networks.
Note that the postprocessing function $\lambda$ plays an important role in how the measurement of the observable in Eq.~\eqref{eq:otheta} is physically realized.
Altogether, this efficiently implements the family of linear classifiers -- which we refer to as \emph{explicit quantum linear classifiers} -- given by
\begin{align}
\label{eq:c_explicit}
    \mathcal{C}^{\text{explicit}}_{\text{qlin}} = \Big\{c_{\mathcal{O}^\lambda_{\theta}, d}(x)=\sign{\tr{\rho_\Phi(x)\mathcal{O}^\lambda_\theta}-d} \text{ }\Big|\text{ } \mathcal{O}^\lambda_\theta\text{ as in Equation~\eqref{eq:otheta}},\text{ }d \in \R \Big\}.
\end{align}

The power of this model lies in the efficient parameterization of the manifold (inside the $4^n$-dimensional vector space of Hermitian operators on $\C^{2^n}$) realized by the quantum feature map together with the parameterized separating hyperplanes that can be attained by $W(\theta)$ and $\lambda$. 
Here also lies a restriction of the explicit quantum linear classifier compared to standard linear classifiers, as in the latter all hyperplanes are possible and in the former only the hyperplanes that lie in the manifold parameterized by $W(\theta)$ and $\lambda$ are possible.
Furthermore, explicit quantum linear classifiers can likely not be efficiently evaluated classically, as computing expectation values $\tr{\rho_\Phi(x)\mathcal{O}^\lambda_\theta}$ is classically intractable for sufficiently complex feature maps and observables~\cite{terhal:advantage_svm, bremner:advantage}.

\paragraph{Implicit quantum linear classifier\footnote{Also called the \emph{quantum kernel estimator}~\cite{havlivcek:qsvm}.}}
Another way to implement a linear classifier is by using the so-called \emph{kernel trick}~\cite{scholkopf:kernel}.
In short, this trick involves expressing the normal vector of the separating hyperplane, -- i.e., the observable $\mathcal{O}$ in the case of quantum linear classifiers -- on a set of training examples $\mathcal{D}$ as a linear combination of feature vectors, resulting in the expression
\[
\mathcal{O}_{\alpha} = \sum_{x' \in \mathcal{D}} \alpha_{x'}\rho_{\Phi(x')} = \sum_{x' \in \mathcal{D}} \alpha_{x'}\ket{\Phi(x')}\bra{\Phi(x')}.
\]
Using this expression we can rewrite the corresponding quantum linear classifier as
\[
c_{\mathcal{O}_\alpha, d}(x) = \sign{\tr{\rho_\Phi(x)\mathcal{O}_\alpha} - d} = \sign{\sum_{x'\in\mathcal{D}}\alpha_{x'} \tr{\rho_{\Phi}(x)\rho_{\Phi}(x')} - d}.
\]
These type of linear classifiers can also be efficiently realized using parameterized quantum circuits.
Using quantum protocols such as the SWAP-test or the Hadamard-test it is possible to efficiently evaluate the overlaps $\Tr[\rho_\Phi(x)\rho_\Phi(x')]$ for the feature map defined in Equation~\eqref{eq:featuremap}.
Afterwards, the optimal parameters $\{\alpha_{x'}\}_{x' \in \mathcal{D}}$ are obtained on a classical computer, e.g., by solving a quadratic program. 
Altogether, this allows us to efficiently implement the family of linear classifiers -- which we refer to as \emph{implicit quantum linear classifiers} -- given by
\begin{align}
\label{eq:c_implicit}
    \mathcal{C}^{\text{implicit}}_{{\text{qlin}}} = \Big\{c_{\mathcal{O}_{\alpha}, d}(x)=\sign{\tr{\rho_\Phi(x)\mathcal{O}_{\alpha}}-d} \text{ }\Big|\text{ }  \mathcal{O}_{\alpha}  = \sum_{x' \in \mathcal{D}}\alpha_{x'}\rho_{\Phi}(x'),\text{ } \alpha \in \mathbb{R}^{|\mathcal{D}|},\text{ } d \in \R \Big\}.
\end{align}

The power of this model comes from the fact that evaluating the overlaps $\tr{\rho_\Phi(x)\rho_\Phi(x')}$ is likely classically intractable for sufficiently complex feature maps~\cite{havlivcek:qsvm}, demonstrating that classical computers can likely neither train nor evaluate this quantum linear classifier efficiently.
Moreover, any quantum linear classifier that is the minimizer of a loss functions that includes \emph{regularization} of the Frobenius norm of the observable can be expressed as an implicit quantum linear classifier~\cite{schuld:kernel}.
However, as we indicate later in Section~\ref{subsec:srm_practice}, this does not mean that we can forego explicit quantum linear classifiers entirely, as in the explicit approach there are unique types of meaningful regularization for which there is no straightforwards correspondence to the implicit approach.

\subsection{Structural risk minimization:\ generalization bounds and model selection}
\label{subsec:srm}

When looking for the optimal family of classifiers for a given learning problem, it is important to carefully select the family's \emph{complexity} (also known as expressivity or capacity).
For instance, in the case of linear classifiers, it is important to select what kind of hyperplanes one allows the classifier to use.
Generally, the more complex the family is, the lower the training errors will be. 
However, if the family becomes overly complex, then it becomes more prone to worse generalization performance (i.e., due to overfitting).
Structural risk minimization is a concrete method that balances this trade-off in order to obtain the best possible performance on unseen examples.
Specifically, structural risk minimization aims to saturate well-established upper bounds on the expected error of the classifier that consist of the sum of two inversely related terms: a \emph{training error} term, and a \emph{complexity term} penalizing more complex models.

In statistical learning theory it is generally assumed that the data is sampled according to some underlying probability distribution $P$ on $\mathcal{X} \times \{-1, +1\}$.
The goal is to find a classifier that minimizes the probability that a random pair sampled according to $P$ is misclassified.
That is, the goal is to find a classifier $c_{f, d}(x) = \mathrm{sign}(f(x) - d)$ that minimize the \emph{expected error} given by 
\begin{align}
\label{eq:generalization_def}    
\text{er}_P(c_{f, d}) = \underset{(x, y) \sim P}{\mathrm{Pr}}\big(c_{f, d}(x) \neq y\big).
\end{align}
As one generally only has access to training examples $\mathcal{D}= \big\{(x_1, y_1), \dots, (x_m, y_m)\big\}$ that are sampled according to the distribution $P$, it is not possible to compute $\text{er}_P$ directly.
Nonetheless, one can try to approximate Equation~\eqref{eq:generalization_def} using \emph{training errors} such as
\begin{align}
\label{eq:empirical_1}
\widehat{\text{er}}_{\mathcal{D}}(c_{f, d}) = \frac{1}{m}\Big|\big\{i \text{ }\big|\text{ } c_{f, d}(x_i) \neq y_i\big\}\Big|,
\end{align}
\begin{align}
\label{eq:empirical_2}
\widehat{\text{er}}^\gamma_{\mathcal{D}}(c_{f, d}) = \frac{1}{m}\Big|\big\{i \text{ }\big|\text{ } y_i\cdot \big(f(x_i) - d\big) < \gamma \big\}\Big|, \enskip\gamma \in \mathbb{R}_{\geq 0}.
\end{align}
Intuitively, $\widehat{\text{er}}_\mathcal{D}$ in Equation~\eqref{eq:empirical_1} represents the frequency of misclassified training examples, and  $\widehat{\text{er}}^\gamma_\mathcal{D}$ in Equation~\eqref{eq:empirical_2} represents the frequency of training examples that are either misclassified or are ``within margin $\gamma$ from being misclassified''.
In particular, for $\gamma = 0$ both training error estimates are identical (i.e., $\widehat{\text{er}}_\mathcal{D}= \widehat{\text{er}}^0_\mathcal{D}$).
When selecting the optimal classifiers from a given model one typically searches for the classifier that minimizes the training error (in practice more elaborate and smooth loss functions are used), which is referred to as \emph{empirical risk minimization}.
The problem that structural risk minimization aims to tackle is how to optimally select a model such that one will have some guarantee that the training error will be close to the expected error.

Structural risk minimization uses expected error bounds  -- two of which we will discuss shortly -- that involve a training error term, and a complexity term that penalizes more complex models.
This complexity term usually scales with a certain measure of the complexity of the family of classifiers.
A well known example of such a complexity measure is the Vapnik-Chervonenkis dimension.
\begin{definition}[VC dimension~\cite{vapnik:dimension}]
\label{def:vc}
    Let $\mathcal{C}$ be a family of functions on $\mathcal{X}$ taking values in $\{-1, +1\}$. 
    We say that a set of points $X = \{x_1, \dots, x_m\} \subset \mathcal{X}$ is shattered by $\mathcal{C}$ if for all $y \in \{-1,+1\}^{m}$, there exists a classifier $c_y\in\mathcal{C}$ that satisfies $c_y(x_i) = y_i$.
    The VC dimension of $\mathcal{C}$ defined as
    \[
        \mathrm{VC}\big(\mathcal{C}\big) = \max \big\{m \text{ }|\text{ } \exists\{x_1, \dots, x_m\}\subset \mathcal{X}\text{ that is shattered by }\mathcal{C}\big\}.
    \]
\end{definition}
Besides the VC dimension we also consider a complexity measure called the \emph{fat-shattering dimension}, which can be viewed as a generalization of the VC dimension to real-valued functions.
An important difference between the VC dimension and the fat-shattering dimension is that the latter also takes into account the so-called \emph{margins} that the family of classifiers can achieve.
Here the margin of a classifier $c_{f, d}(x) = \sign{f(x) - d}$ on a set of examples $\{x_i\}_{i=1}^m$ is given by $\min_i |f(x_i) - d|$.
Throughout the literature, this is often referred to as the functional margin.

\begin{definition}[Fat-shattering dimension~\cite{kearns:fs}]
\label{def:fs}
    Let $\mathcal{F}$ be a family of real-valued functions on $\mathcal{X}$. 
    We say that a set of points $X = \{x_1, \dots, x_m\} \subset \mathcal{X}$ is $\gamma$-shattered by $\mathcal{F}$ if there exists an $s \in \mathbb{R}^{m}$ such that for all $y \in \{-1,+1\}^{m}$, there exists a function $f_y\in\mathcal{F}$ satisfying
    \begin{align*}
        f_y(x_i) \begin{cases}
        \leq s_i-\gamma & \text{ if } y_i=-1,\\
        \geq s_i+\gamma & \text{ if } y_i=+1.
        \end{cases}
    \end{align*}
    The fat-shattering dimension of $\mathcal{F}$ is a function $\fat_\mathcal{F} : \mathbb{R} \rightarrow \mathbb{Z}_{\geq 0}$ that maps
    \[
        \fat_\mathcal{F}(\gamma) = \max\big\{m \text{ }\big|\text{ } \exists \{x_1, \dots, x_m\} \subset \mathcal{X}\text{ that is $\gamma$-shattered by $\mathcal{F}$}\big\}.
    \]
\end{definition}

We will now state two expected error bounds that can be used to perform structural risk minimization. 
These error bounds theoretically quantify how an increase in model complexity (i.e., VC dimension or fat-shattering dimension) results in a worse expected error (i.e., due to overfitting).
First, we state the expected error bound that involves the VC dimension.
\begin{theorem}[Expected error bound using VC dimension~\cite{wolf:lecture_notes}]
\label{thm:expected_error_vc}
    Consider a set of functions $\mathcal{C}$ on $\mathcal{X}$ taking values in $\{-1, +1\}$. Suppose $\mathcal{D}= \{(x_1, y_1), \dots, (x_m, y_m)\}$ is sampled using $m$ independent draws from $P$. Then, with probability at least $1-\delta$, the following holds for all $c \in \mathcal{C}$: 
    \begin{align}
    \label{eq:error_bound_vc}
        \mathrm{er}_P(c) \leq \widehat{\mathrm{er}}_\mathcal{D}(c) + 62\sqrt{\frac{k}{m}} + 3\sqrt{\frac{\log(2/\delta)}{2m}}
    \end{align}
    where $k = \mathrm{VC}\big(\mathcal{C}\big)$.
\end{theorem}

Next, we state the expected error bound that involves the fat-shattering dimension.
One possible advantage of using the fat-shattering dimension instead of the VC dimension is that it can take into account the margin that the classifier achieves on the training examples.
This turns out to be useful since this margin can be used to more precisely fine-tune the expected error bound.

\begin{theorem}[Expected error bound using fat-shattering dimension~\cite{bartlett:generalization1}]
\label{thm:expected_error_fs}
    Consider a set of real-valued functions $\mathcal{F}$ on $\mathcal{X}$. Suppose $\mathcal{D}= \{(x_1, y_1), \dots, (x_m, y_m)\}$ is sampled using $m$ independent draws from $P$. Then, with probability at least $1-\delta$, the following holds for all $c(x) = \sign{f(x) - d}$ with $f \in \mathcal{F}$ and $d \in \mathbb{R}$:
    \begin{align}
    \label{eq:error_bound_fs}
        \mathrm{er}_P(c) \leq \widehat{\mathrm{er}}^\gamma_\mathcal{D}(c) + \sqrt{\frac{2}{m}\Big(k \log(34em/k)\log_2(578m) + \log(4/\delta)\Big)}. 
    \end{align}
    where $k=\fat_\mathcal{F}(\gamma/16)$.
\end{theorem}
\begin{remark*}
If the classifier can correctly classify all examples in $\mathcal{D}$, then the optimal choice of~$\gamma$ in the above theorem is the margin achieved on the examples in $\mathcal{D}$, i.e., $\gamma=\min_{x_i \in \mathcal{D}}\big|f(x_i) - d\big|$.
\end{remark*}

Generally, the more complex a family of classifiers is, the larger its generalization errors are.
This correlation between a family's complexity and its generalization errors is theoretically quantified in Theorems~\ref{thm:expected_error_vc} and~\ref{thm:expected_error_fs}.
Specifically, the more complex the family is the larger its VC dimension will be, which strictly increases the second term in Equation~\ref{eq:error_bound_vc} that corresponds to the generalization error.
Note that for the fat-shattering dimension in Theorem~\ref{thm:expected_error_fs} this is not as obvious.
In particular, a more complex model could achieve a larger margin $\gamma$, which actually decreases the second term in Equation~\ref{eq:error_bound_fs} that corresponds to the generalization error.

\begin{figure*}[h!]
\centering
\includegraphics[width=0.6\textwidth]{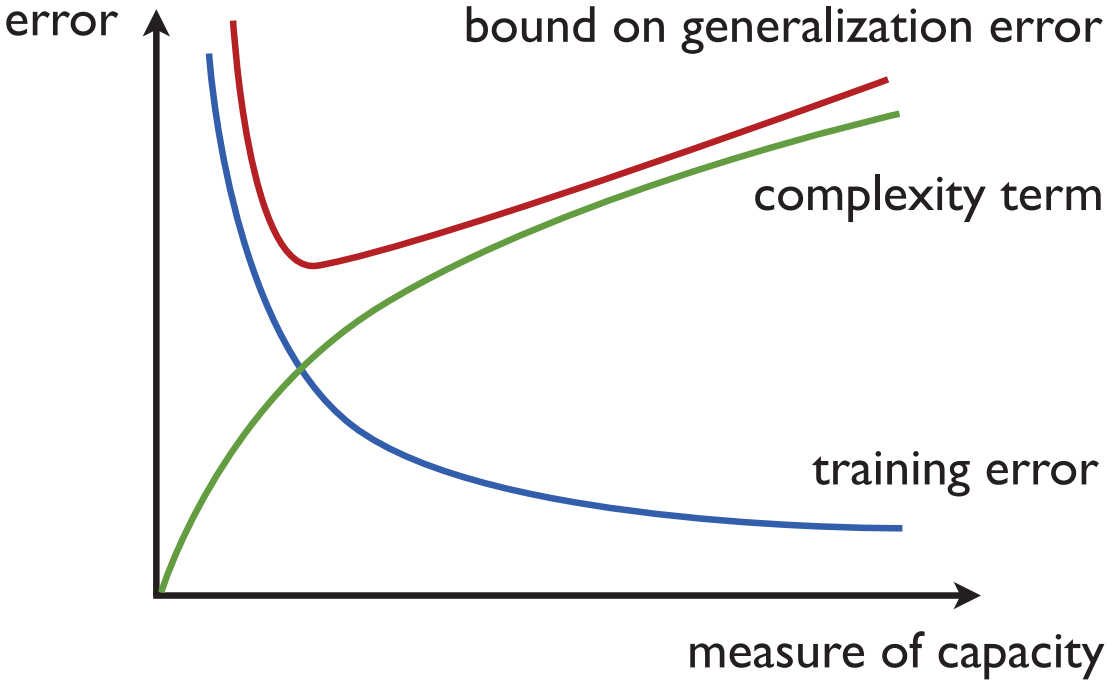}
\caption{\label{fig:srm}Illustration of structural risk minimization taken from~\cite{mohri:foundations}.
Increasing the complexity of the classifier family causes the training error (blue) to decrease, while it increases the complexity term (green).
Structural risk minimization selects the classifier minimizing the expected error bound in Eqs.~\eqref{eq:error_bound_vc} and~\eqref{eq:error_bound_fs} given by the sum of the training error and the complexity term (red).}
\end{figure*}

Theorems~\ref{thm:expected_error_vc} and~\ref{thm:expected_error_fs} establish that in order to minimize the expected error, we should aim to minimize either of the sums on the right-hand side of Equations~\eqref{eq:error_bound_vc} or~\eqref{eq:error_bound_fs} (depending on which complexity measure one wishes to focus on).
Note that in both cases the first term corresponds to a training error and the second term corresponds to a complexity term that penalizes more complex models.
Crucially, the effect that the complexity measure of the family of classifiers has on these terms is inversely related.
Namely, a large complexity measure generally gives rise to smaller training errors, but at the cost of a larger complexity term.
Balancing this trade-off is precisely the idea behind structural risk minimization.
More precisely, structural risk minimization selects a classifier that minimizes either of the expected error bounds stated in Theorem~\ref{thm:expected_error_vc} or~\ref{thm:expected_error_fs}, by selecting the classifier from a family whose complexity measure is fine-tuned in order to balance both terms on the right-hand side of Equations~\eqref{eq:error_bound_vc} or~\eqref{eq:error_bound_fs}.
Note that limiting the VC dimension and fat-shattering dimension does not achieve the same theoretical guarantees on the generalization error, and it will generally give rise to different performances in practice (as also discussed Section~\ref{subsec:expressivity}).
An overview of the trade-off in the error bounds stated in Theorems~\ref{thm:expected_error_vc} and~\ref{thm:expected_error_fs} is depicted in Figure~\ref{fig:srm}.

\section{Structural risk minimization for quantum linear classifiers}
\label{sec:srm_for_qlc}

In this section we theoretically analyze and quantify the influence that model parameters of quantum linear classifiers have on the trade-off in structural risk minimization. 
We first analyze the effect that model parameters have on the complexity term (i.e., the green line in Figure~\ref{fig:srm}) and afterwards we analyze their effect on the training error (i.e., the blue line in Figure~\ref{fig:srm}).
Specifically, in Section~\ref{subsec:fs-dim} we analyze the complexity term by establishing analytic upper bounds on complexity measures (i.e., the VC dimension and fat-shattering dimension) of quantum linear classifiers.
In Section~\ref{subsec:expressivity} we study the influence that model parameters which influence the established complexity measure bounds have on the training error term.
Finally, in Section~\ref{subsec:srm_practice}, we discuss how to implement structural risk minimization of quantum linear classifiers based on the obtained results.

\subsection{Complexity of quantum linear classifiers: fat-shattering and VC dimension}
\label{subsec:fs-dim}

In this section we determine the two complexity measures defined in the previous section -- i.e., the fat-shattering dimension and VC dimension -- for families of quantum linear classifiers. 
As a result, we identify model parameters that allow us to control the complexity term in the expected error bounds of Theorems~\ref{thm:expected_error_vc} and~\ref{thm:expected_error_fs}.
In particular, these model parameters can therefore be used to balance the trade-off considered by structural risk minimization, as depicted in Figure~\ref{fig:srm}.
Throughout this section we fix the feature map to be the one defined Equation~\eqref{eq:featuremap} and we allow our separating hyperplanes to come from a family of observables $\mathbb{O} \subseteq \mathrm{Herm}\left(\mathbb{C}^{2^n} \right)$ (e.g., the family of observables implementable using either the explicit or implicit realization of quantum linear classifiers).
Our goal is to determine analytical upper bounds on complexity measures of the resulting family of quantum linear classifiers.

First, we show that the VC dimension of a family of quantum linear classifiers is upper bounded by the dimension of the span of the observables that it uses. 
This in turn is upper bounded by the square of the dimension of the space upon which the observables act nontrivially.  
We remark that while the VC dimension of quantum linear classifiers also has a clear dependence on the feature map, we chose to focus on the observables because the resulting upper bounds give rise to more explicit guidelines on how to tune the quantum model to perform structural risk minimization (as we discuss in more detail in Section~\ref{subsec:srm_practice}).
We defer the proof to Appendix~\ref{appendix:vc}.

\begin{restatable}{proposition}{vcdim}
\label{thm:vc}
Let $\mathbb{O}\subseteq \mathrm{Herm}\left(\mathbb{C}^{2^n} \right)$ be a family of $n$-qubit observables with $r = \dim\big(\sum_{\mathcal{O} \in \mathbb{O}}\mathrm{Im}\hspace{1pt}\mathcal{O}\big)\footnote{Here $\sum$ denotes the sum of vector spaces and $\mathrm{Im}\hspace{1pt}\mathcal{O}$ denotes the image (or column space) of the operator $\mathcal{O}$}.$.
Then, the VC dimension of 
\begin{align}
\label{eq:c_qlin_O}
\mathcal{C}^{\mathbb{O}}_{\mathrm{qlin}} = \Big\{c(x) = \sign{\tr{\mathcal{O}\rho_\Phi(x)} - d}\text{ }\big|\text{ }\mathcal{O}\in \mathbb{O}\text{, }d \in \mathbb{R}\Big\}
\end{align}
satisfies
\begin{align}
\label{eq:vc_qlin}
    \mathrm{VC}\big(\mathcal{C}^{\mathbb{O}}_{\mathrm{qlin}}\big) \leq \dim\big(\mathrm{Span}\big(\mathbb{O}\big) \big) + 1 \leq r^2 + 1.
\end{align}
\end{restatable}
\begin{remark*}
The quantity $r$ in the above proposition is related to the ranks of the observables.
Specifically, note that for any two observables $\mathcal{O}, \mathcal{O}' \in \mathrm{Herm}\big(\C^{2^n}\big)$ we have that
\[
\dim\big(\mathrm{Im}\hspace{1pt}\mathcal{O} + \mathrm{Im}\hspace{1pt}\mathcal{O}' \big) = \mathrm{rank}\big(\mathcal{O}\big) + \mathrm{rank}\big(\mathcal{O}'\big) - \mathrm{dim}\big(\mathrm{Im}\hspace{1pt}\mathcal{O} \cap \mathrm{Im}\hspace{1pt}\mathcal{O}'\big).
\]
\end{remark*}

The above proposition implies the (essentially obvious) result that VC dimension of a family of implicit quantum linear classifiers is upper bounded by the number of training examples (i.e., the operators $\{\rho_\Phi(x)\}_{x \in \mathcal{D}}$ span a subspace of dimension at most $\big|\mathcal{D}\big|$).
We are however more interested in the application of the above proposition to explicit quantum linear classifiers.
In this case, we choose to focus on the upper bound $r^2 + 1$ because it has interpretational advantages as to what parts of the model one has to tune from the perspective of structural risk minimization (i.e., recall from Section~\ref{sec:srm_for_qlc} that one way to perform structural risk minimization is to tune the VC dimension). 
Moreover, in the case of explicit quantum linear classifiers, the bound $r^2 + 1$ is only quadratically worse than the bound $\dim\big(\mathrm{Span}\big(\mathbb{O}\big)\big) + 1$.
To see this, we consider a family of explicit quantum linear classifiers with observables $\mathbb{O}_\mathrm{explicit} = \big\{\mathcal{O}_\theta^\lambda\big\}$, where
\begin{align*}
    \mathcal{O}_\theta^\lambda  = W^\dagger(\theta) \cdot \mathrm{diag}\big(\lambda(0), \dots, \lambda(2^n-1) \big)\cdot W(\theta)
\end{align*}
and we denote $W(\theta)\ket{i} = \ket{\psi_i(\theta)}$.
Next, suppose that $\lambda(j) = 0$ for all $j > L$ and define 
\begin{align}
\label{eq:vectorspace1}
    H &= \mathrm{Span}_\mathbb{C}\Big\{\ket{\psi_0(\theta)}, \dots, \ket{\psi_{L}(\theta)}\text{ }:\text{ }\theta \in \mathbb{R}^m\Big\},\\
    \label{eq:vectorspace2}
    V &= \mathrm{Span}_\mathbb{R}\Big\{\sum_{i=0}^L\lambda(i)\ket{\psi_i(\theta)}\bra{\psi_i(\theta)}\text{ }:\text{ }\theta \in \mathbb{R}^m\Big\},
\end{align}
Then, Proposition~\ref{thm:vc} states that
\begin{align*}
\mathrm{VC}\big(\mathcal{C}^{\mathbb{O}_{\mathrm{explicit}}}_{\mathrm{qlin}}\big) \leq \dim\big(V\big) + 1 \leq \dim(H)^2 + 1.
\end{align*}
Now, by the following lemma, we indeed find that the bound $r^2 + 1$ is only quadratically worse than the bound $\dim\big(\mathrm{Span}\big(\mathbb{O}\big)\big) + 1$. 
We again defer the proof to Apppendix~\ref{appendix:vc}.
\begin{restatable}{lemma}{quadratic}
\label{lemma:quadratic}
The vector spaces defined in Eq.~\eqref{eq:vectorspace1} and Eq.~\eqref{eq:vectorspace2} satisfy\footnote{Note that there exists ansatzes for which the inequalities are strict, i.e., $\dim(H) < \dim(V) < \dim(H)^2$ (e.g., see the first example discussed in Section~\ref{subsec:srm_practice}).}
\[
\dim(H) \leq \dim(V) \leq \dim(H)^2.
\]
\end{restatable}

Therefore, if we sufficiently limit $r =\dim (H)$, then this also limits $\dim\big(\mathrm{Span}\big(\mathbb{O}\big)\big) = \dim(V)$.
Moreover, even though $\dim\big(\mathrm{Span}\big(\mathbb{O}\big)\big) + 1$ can provide a tighter bound, it can still be advantageous to study the bound $r^2 + 1$ because it might have interpretational advantages. 
Specifically, it might be easier to construct cases of ansatze where the latter bound allows us to identify a controlable hyperparameter that controls the VC dimension (as we discuss in more detail in Section~\ref{subsec:srm_practice}).

Note that the quantity $r$ defined in the above proposition, depends on both the structure of the ansatz $W$ as well as the post-processing function $\lambda$.
One way to potentially limit $r$ is by varying the rank of the final measurement (i.e., the value $L$ defined above).
However, for several ansatzes in literature, having either a low-rank or a high-rank final measurement will not make a difference in terms of the VC dimension bound $r^2 + 1$\footnote{The relationship between the quantity $r$ and the ranks of the observable can be made explicit by considering the overlaps between the images of the observables. A more detailed explanation of this can be found in Appendix~\ref{appendix:rvsr}.}.
To see this, consider an ansatz consisting of a single layer of parameterized $X$-rotations on all qubits, where each rotation is given a separate parameter.
Already for this simple ansatz even the first columns $\{\bigotimes_{i=1}^n X_i(\theta_i)\ket{0} \mid \theta \in [0, 2\pi)^n\}$ span the entire $n$-qubit Hilbert space.
In particular, the above proposition gives the same VC dimension upper bound for the cases where the final measurement is of rank $L=1$, and where it is of full rank $L = 2^n$ (i.e., we have no guarantee that limiting $L$ limits the VC dimension).
This motivates us to design ansatzes for which subsets of columns do not span the entire Hilbert space when varying the variational parameter $\theta$.
On the other hand, to exploit the bound $\dim\big(\mathrm{Span}\big(\mathbb{O}\big)\big) + 1$ one needs to consider the span of the projectors onto the first $L$ columns in the vector space of Hermitian operators.
This quantity can be slightly less intuitive than the span of the first $L$ columns in the $n$-qubit Hilbert space, and in Section~\ref{subsec:srm_practice} we show that this latter quantity can already be used to affirm the effectiveness of certain regularization techniques.
Specifically, in Section~\ref{subsec:srm_practice} we discuss examples of ansatzes for which subsets of columns do not span the entire Hilbert space when varying the variational parameter, and we explain how they allow for structural risk minimization by limiting the rank of the final measurement.

Next, we show that the fat-shattering dimension of a family of quantum linear classifiers is related to the Frobenius norm of the observables that it uses.
In particular, we show that we can control the fat-shattering dimension of a family of quantum linear classifiers by limiting the Frobenius norm of its observables.
We defer the proof to Appendix~\ref{appendix:fs}, where we also discuss the implications of this result in the probably approximately correct (PAC) learning framework.

\begin{restatable}{proposition}{fsdimupperbound}
\label{thm:fs_upperbound}
Let $\mathbb{O}\subseteq \mathrm{Herm}\left(\mathbb{C}^{2^n} \right)$ be a family of $n$-qubit observables with $\eta = \max_{\mathcal{O}\in \mathbb{O}} \|\mathcal{O}\|_F$. 
Then, the fat-shattering dimension of 
\begin{align}
\label{eq:f_qlin_O}
    \mathcal{F}^{\mathbb{O}}_{\mathrm{qlin}} = \Big\{f_{\mathcal{O}, d}(x) = \tr{\mathcal{O} \rho_\Phi(x)} - d \text{ }\big|\text{ } \mathcal{O} \in \mathbb{O}, \text{ }d \in \mathbb{R}\Big\} 
\end{align}
is upper bounded by
\begin{align}
\label{eq:fs_qlin}
    \fat_{\mathcal{F}^{\mathbb{O}}_{\mathrm{qlin}}}(\gamma)\leq O\left(\frac{\eta^2}{\gamma^2}\right).
\end{align}
\end{restatable}
\begin{remark*}
The upper bound in the above proposition matches the result discussed in~\cite{liu:power_appendix}.
This was derived independently by one of the authors of this paper~\cite{vreumingen:msc}, and we include it here for completeness.
\end{remark*}

The above proposition shows that the fat-shattering dimension of a family of explicit quantum linear classifiers can be controlled by limiting $||\mathcal{O}_\theta^\lambda||_F = \sqrt{\sum_{i = 1}^{2^n}\lambda(i)^2}$.
In particular, it shows that the selection of the postprocessing function $\lambda$ is important when tuning the complexity of the family ofr classifiers. 
Furthermore, the above proposition shows that the fat-shattering dimension of a family of implicit quantum linear classifiers can be controlled by limiting $||\mathcal{O}_{\alpha}||_F \leq ||\alpha||_1$.
It is important to note that the Frobenius norm itself does not fully characterize the generalization performance of a family of quantum linear classifiers.
Specifically, plugging Theorem~\ref{thm:fs_upperbound} into Proposition~\ref{thm:expected_error_fs} we find that the generalization performance bounds depend on both the Frobenius norm as well as the functional margin on training examples\footnote{Recall that the functional margin of $c_{f, d}(x) = \sign{f(x) - d}$ on a set of examples $\{x_i\}$ is $\min_i | f(x_i) - d |$.}.
Therefore, to optimize the generalization performance bounds one has to minimize the Frobenius norm, while ensuring the functional margin on training examples stays large.
Note that one way to achieve this is by maximizing the so-called geometric margin, which on a set of example $\{x_i\}$ is given by $\min_i \big|\tr{\mathcal{O}\rho_\Phi(x_i)} - d\big| / ||\mathcal{O}||_F$.

\subsection{Expressivity of quantum linear classifiers:\ model parameters \& errors}
\label{subsec:expressivity}

Having established that the quantity $r$ defined in Proposition~\ref{thm:vc} and the Frobenius norms of the observables influence the complexity of the family of quantum linear classifiers (i.e., the green line in Figure~\ref{fig:srm}), we will now study the influence of these parameters on the training errors that the classifiers can achieve (i.e., the blue line in Figure~\ref{fig:srm}).
First, we study the influence of these model parameters on the ability of the classifiers to correctly classify certain sets of examples.
Afterwards, we study the influence of these model parameters on the margins that the classifiers can achieve.

Recall from the previous section that the VC dimension of certain families of quantum linear classifiers depends on the rank of the observables that it uses.
For instance, if the observables are such that their images are (largely) overlapping, then the quantity $r$ defined in Proposition~\ref{thm:vc} can be controlled by limiting the ranks of all observables.
In Section~\ref{subsec:srm_practice} we use this observation to construct ansatzes for which the VC dimension bound can be tuned by varying the rank of the observable measured on the output of the circuit.
Since the VC dimension is only concerned with whether an example is correctly classified (and not what margin it achieves), we choose to investigate the influence of the rank on being able to correctly classify certain sets of examples.
In particular, we show that any set of examples that can be correctly classified using a low-rank observable, can also be correctly classified using a high-rank observable.
Moreover, we also show that there exist sets of examples that can only be correctly classified using observables of at least a certain rank.
We defer the proof to Appendix~\ref{appendix:rank}.

\begin{restatable}{proposition}{zeromargin}
\label{prop:expressivity_zeromargin}
Let $\mathcal{C}_{\mathrm{qlin}}^{(r)}$ denote the family of quantum linear classifiers corresponding to observables of exactly rank $r$, that is,
\begin{align}
    \label{eq:c_qlin_r}
    \mathcal{C}_{\mathrm{qlin}}^{(r)} = \Big\{c(\rho) = \sign{\tr{\mathcal{O}\rho} - d}\text{ }\big|\text{ }\mathcal{O} \in \mathrm{Herm}\big(\C^{2^n}\big)\text{, }\mathrm{rank}\big(\mathcal{O}\big) = r\text{, }d \in \R \Big\}
\end{align}
Then, the following statements hold:
\begin{enumerate}[label=(\roman*)] 
    \item For every finite set of examples $\mathcal{D}$ that is correctly classified by a quantum linear classifier $c \in \mathcal{C}_{\mathrm{qlin}}^{(k)}$ with $0 < k < 2^n$, there exists a quantum linear classifier $c \in \mathcal{C}_{\mathrm{qlin}}^{(r)}$ with $r > k$ that also correctly classifies $\mathcal{D}$.
    \item There exists a finite set of examples that can be correctly classified by a classifier $c \in \mathcal{C}^{(r)}_{\mathrm{qlin}}$, but which no classifier $c' \in \mathcal{C}^{(k)}_{\mathrm{qlin}}$ with $k < r$ can classify correctly.
\end{enumerate}
\end{restatable}

Note that in the above proposition we define our classifiers in such a way that high-rank classifiers do not subsume low-rank classifiers.
In particular, the family of observables that $\mathcal{C}_{\mathrm{qlin}}^{(r)}$ and $\mathcal{C}_{\mathrm{qlin}}^{(k)}$ use are completely disjoint for $k \neq r$.
The construction behind the proof of the above proposition is inspired by tomography of observables.
Specifically, we construct a protocol that queries a quantum linear classifier and based on the assigned labels checks whether the underlying observable is approximately equal to a fixed target observable of a certain rank. 
In particular, we can use this to test whether the underlying observable is really of a given rank, as no low-rank observable can agree with a high-rank observable on the assigned labels during this protocol.
Note that if we could query the expectation values of the observable, then tomography would be straightforward.
However, the classifier only outputs the sign of the expectation value, which introduces a technical problem that we circumvent.
Our protocol could be generalized to a more complete tomographic-protocol which uses queries to a quantum linear classifier in order to find the spectrum of the underlying observable.

Next, we investigate the effect that limitations of the rank of the observables used by a family of quantum linear classifier have on its ability to implement certain families of standard linear classifiers.
In particular, assuming that the feature map is bounded (i.e., all feature vectors have finite norm), then the following proposition establishes the following chain of inclusions:
\begin{align}
    \label{eq:chain}
    \mathcal{C}_{\mathrm{lin}} \text{ on $\R^{2^n}$} &\subseteq \mathcal{C}^{(\leq 1)}_{\mathrm{qlin}} \text{ on $n+1$ qubits} \subseteq \dots \subseteq \mathcal{C}^{(\leq r)}_{\mathrm{qlin}} \text{ on $n+1$ qubits} \subseteq \dots \subseteq   \mathcal{C}_{\mathrm{lin}} \text{ on $\R^{4^n}$},
\end{align}
where $\mathcal{C}^{(\leq r)}_{\mathrm{qlin}}$ denotes the family of quantum linear classifiers using observables of rank at most $r$.
Note that $\mathcal{C}^{(\leq r)}_{\mathrm{qlin}} \subsetneq \mathcal{C}^{(\leq r+1)}_{\mathrm{qlin}}$ is strict due to Proposition~\ref{prop:expressivity_zeromargin}.
We defer the proof to Appendix~\ref{appendix:comparisson}.

\begin{restatable}{proposition}{comparison}
\label{prop:comparison}
Let $\mathcal{C}_{\mathrm{lin}}(\Phi)$ denote the family of linear classifiers that is equipped with a feature map $\Phi$.
Also, let $\mathcal{C}^{(\leq r)}_{\mathrm{qlin}}(\Phi')$ denote the family of quantum linear classifiers that uses observables of rank at most $r$ and which is equipped with a quantum feature map $\Phi'$.
Then, the following statements hold:
\begin{enumerate}[label=(\roman*)]
    \item For every feature map $\Phi: \R^{\ell} \rightarrow \R^N$ with $\sup_{x \in \R^\ell}||\Phi(x)|| = M < \infty$, there exists a feature map $\Phi': \R^{\ell} \rightarrow \R^{N+1}$ such that $||\Phi'(x)||= 1$ for all $x \in \R^\ell$ and the families of linear classifiers satisfy $\mathcal{C}_{\mathrm{lin}}(\Phi) \subseteq \mathcal{C}_{\mathrm{lin}}(\Phi')$.
    \item For every feature map $\Phi: \R^{\ell} \rightarrow \R^N$ with $||\Phi(x)|| = 1$ for all $x \in \R^\ell$, there exists a quantum feature map $\Phi': \R^{\ell} \rightarrow \mathrm{Herm}\left(\C^{2^n}\right)$ that uses $n = \lceil\log N +1 \rceil + 1$ qubits such that the families of linear classifiers satisfy $\mathcal{C}_{\mathrm{lin}}(\Phi) \subseteq \mathcal{C}^{(\leq 1)}_{\mathrm{qlin}}(\Phi')$.
    \item For every quantum feature map $\Phi:\R^{\ell} \rightarrow \mathrm{Herm}\left(\C^{2^n}\right)$, there exists a classical feature map $\Phi':\R^{\ell} \rightarrow \R^{4^n}$ such that the families of linear classifiers satisfy $\mathcal{C}_{\mathrm{qlin}}(\Phi) = \mathcal{C}_{\mathrm{lin}}(\Phi')$.
\end{enumerate}
\end{restatable}

Recall from the previous section that the fat-shattering dimension of a family of linear classifiers depends on the Frobenius norm of the observables that is uses.
In the following proposition we show that tuning the Frobenius norm changes the margins that the model can achieve, which gives rise to better generalization performance (as discussed in Section~\ref{subsec:srm}).
In particular, we show that there exist sets of examples that can only be classified with a certain margin by a classifier that uses an observable of at least a certain Frobenius norm.
We defer the proof to Appendix~\ref{appendix:fs_expr}.

\begin{restatable}{proposition}{nonzeromargin}
\label{prop:expressivity_nonzeromargin}
Let $\mathcal{C}^{(\eta)}_{\mathrm{qlin}}$ denote the family of quantum linear classifiers corresponding to all $n$-qubit observables of Frobenius norm $\eta$, that is,
\begin{align}
    \label{eq:c_qlin_eta}
    \mathcal{C}_{\mathrm{qlin}}^{(\eta)} = \Big\{c(\rho) = \sign{\tr{\mathcal{O}\rho} - d}\text{ }\big|\text{ }\mathcal{O} \in \mathrm{Herm}\big(\C^{2^n}\big)\text{ with }||\mathcal{O}||_F = \eta\text{, }d \in \R \Big\}.
\end{align}
Then, for every $\eta \in \mathbb{R}_{>0}$ and $0 < m \leq 2^n$ there exists a set of $m$ examples consisting of binary labeled $n$-qubit pure states that satisfies the following two conditions:
\begin{enumerate}[label=(\roman*)]
    \item There exists a classifier $c \in \mathcal{C}^{( \eta)}_{\mathrm{qlin}}$ that correctly classifies all examples with margin $\eta/\sqrt{m}$.
    \item No classifier $c' \in \mathcal{C}^{( \eta')}_{\mathrm{qlin}}$ with $\eta' < \eta$ can classify all examples correctly with margin $\geq \eta / \sqrt{m}$.
\end{enumerate}
\end{restatable}

In conclusion, in Proposition~\ref{thm:vc} we showed that in certain cases the rank of the observables control the model's complexity (e.g., if the observables have overlapping images), and in Proposition~\ref{prop:expressivity_zeromargin} we showed that the rank also controls the model's ability to achieve small training errors.
Moreover, in Proposition~\ref{prop:expressivity_nonzeromargin} we similarly showed that the Frobenius norm not only controls the model's complexity (see Proposition~\ref{thm:fs_upperbound}), but that it also controls the model's ability to achieve large functional margins.
However, note that tuning each model parameter achieves a different objective.
Namely, increasing the rank of the observable increases the ability to correctly classify sets of examples, whereas increasing the Frobenius norm of the observable increases the margins that it can achieve.
For example, one can increase the Frobenius norm of an observable by multiplying it with a positive scalar which increases the margin it achieves, but in order to correctly classify the sets of examples discussed in Proposition~\ref{prop:expressivity_zeromargin} one actually has to increase the rank of the observable.

\subsection{Structural risk minimization for quantum linear classifiers in practice}
\label{subsec:srm_practice}

Having established how certain model parameters of quantum linear classifiers influence both the model's complexity and its ability to achieve small training errors, we now discuss how to use these results to implement structural risk minimization of quantum linear classifiers in practice.
In particular, we will discuss a common approach to structural risk minimization called \emph{regularization}.
In short, what regularization entails is instead of minimizing only the training error $E_{\mathrm{train}}$, one simultaneously minimizes an extra term $h(\omega)$, where $h$ is a function that takes larger values for model parameters $\omega$ that correspond to more complex models.
In this section, we discuss different types of regularization (i.e., different choices of the function $h$) that can be performed in the context of quantum linear classifiers based on the results of the previous section.
These types of regularization help improve the performance of quantum linear classifiers in practice, without putting more stringent requirements on the quantum hardware and are thus NISQ-suitable.

To illustrate how Proposition~\ref{thm:vc} can be used to implement structural risk minimization in the explicit approach, consider the setting where we have a parameterized quantum circuit $W(\theta)$ (with $\theta \in \mathbb{R}^p$) followed by a fixed measurement that projects onto the first $\ell$ computational basis states.
To use the bound $r^2 + 1$ from Proposition~\ref{thm:vc} one has to compute the quantity 
\begin{align}
\label{eq:bound1}
\dim_{\mathbb{C}} \Big(\mathrm{Span}_\mathbb{C}\big\{\ket{\psi_i(\theta)} \text{ }:\text{ } i = 1, \dots \ell,\text{ }\theta \in \mathbb{R}^p \big\}\Big),
\end{align}
where $\ket{\psi_i(\theta)}$ denotes the $i$th column of $W(\theta)$.
To use the other bound $\dim\big(\mathrm{Span}\big(\mathbb{O}\big)\big) + 1$ from Proposition~\ref{thm:vc} one has to compute the quantity 
\begin{align}
\label{eq:bound2}
\dim_\mathbb{R} \Big(\mathrm{Span}_\mathbb{R}\big\{\sum_{i = 1}^\ell \ket{\psi_i(\theta)}\bra{\psi_i(\theta)} \text{ }:\text{ } \theta \in \mathbb{R}^p \big\}\Big),
\end{align}
Although both are of course possible, in some cases it is slightly easier to see how the quantity in Eq.~\eqref{eq:bound1} scales with respect to $\ell$.
Specifically, utilizing the quantity in Eq.~\eqref{eq:bound1} already leads to interesting ansatze that allow for structural risk minimization by limiting $\ell$.
As discussed below Proposition~\ref{thm:vc}, setting $\ell$ to be either large or small will not influence the upper bound on the VC dimension independently of the structure of the parameterized quantum circuit ansatz $W$.
The proposition therefore motivates the design of ansatzes whose first $\ell$ columns define a manifold when varying the variational parameter that is contained in a relatively low-dimensional linear subspace. 
Specifically, in this case Proposition~\ref{thm:vc} results in nontrivial bounds on the VC dimension that we aim to control by varying $\ell$.
We now give three examples of ansatzes that allow one to control the upper bound on the VC dimension by varying~$\ell$.
In particular, these ansatzes allow structural risk minimization to be implemented by regularizing with respect to the rank of the final measurement.
\begin{itemize}

\item For the first example, split up the qubits up in a ``control register'' of size $c$ and a ``target register'' of size $t$ (i.e., $n = t +c$).
Next, let $C\mathrm{-}U_i(\theta_i)$ denote the controlled gate that applies the $t$-qubit parameterized unitary $U_i(\theta_i)$ to the target register if the control register is in the state $|i\rangle$.
Finally, consider the ansatz
\[
W(\theta) = \big[C\mathrm{-}U_{2^c}(\theta_{2^c})\big]\cdot \ldots \cdot \big[C\mathrm{-}U_1(\theta_1)\big].\footnote{We can control the depth of $W(\theta)$ by either limiting the size of the control register or by simply dropping some of the controlled parameterized unitaries (i.e., setting $U_i(\theta_i) = I$).}
\]
Note that the matrix of $W(\theta)$ is given by the block matrix
\[
W(\theta) = \begin{pmatrix} U_1(\theta_1) &               &        & \\ 
                                          & U_2(\theta_2) &        & \\
                                          &               & \ddots & \\
                                          &               &        & U_{2^c}(\theta_{2^c})\end{pmatrix}.
\]

For this choice of ansatz, if the final measurement projects onto $\ell = m2^t$ ($m < 2^c$) computational basis states, then by Proposition~\ref{thm:vc} the VC dimension is at most $\ell^2 + 1$.
Note that $t$ is a controllable hyperparameter that can be used to tune the VC dimension. 
In particular, we can set it such that the resulting VC dimension is not exponential in $n$.
Let us now consider the other bound $\dim\big(\mathrm{Span}\big(\mathbb{O}\big)\big) + 1$ from Proposition~\ref{thm:vc}.    
For this choice of ansatz, computing the quantity in Eq.~\eqref{eq:bound2} is also straightforward due to the block structure of the unitary.
Moreover, for this choice of ansatz the inequalities in Lemma~\ref{lemma:quadratic} are strict, which shows why being able to compute the quantity in Eq.~\eqref{eq:bound1} does not always imply that we can also compute the quantity in Eq.~\eqref{eq:bound2} (i.e., one is not simply the square of the other).

\item For the second example, consider an ansatz that is composed of parameterized gates of the form $U(\theta) = e^{i\theta P}$ for some Pauli string $P \in \{X, Y, Z, I\}^{\otimes n}$.
Specifically, consider the ansatz
\[
W(\theta) = e^{i\theta_d P_d}\cdot \ldots \cdot e^{i\theta_1 P_1}.
\]
By the bound $r^2 + 1$ from Proposition~\ref{thm:vc}, for this choice of ansatz if the final measurement projects onto $\ell$ computational basis states the VC dimension is at most $r^2 + 1$, where $r = \ell \cdot 2^d$.
This bound is obtained by computing the quantity in Eq.~\eqref{eq:bound1}, which can be done by noting that a column of the unitary $U(\theta)$ spans a subspace of dimension at most~2 when varying the variational parameter~$\theta$.
Moreover, subsequent layers of $U(\theta)$ will only increase the dimension of the span of a column by at most a factor~2.
Thus, when applying $U(\theta)$ a total of $d$ times, the dimension of the span of any $\ell$ columns of $W(\theta)$ is at most $r = \ell\cdot 2 ^d$.
Also in this construction we note that $d$ is a controllable hyperparameter that can be used to tune the VC dimension. 
In particular, we can set it such that the resulting VC dimension is not exponential in $n$.
For this particular choice of ansatze, computing the quantity in Eq.~\eqref{eq:bound2} might also be possible, but it is a bit more involved and not necessary for our main goal of establishing that $\ell$ controls the VC dimension.
In particular, one might be able to compute the quantity in Eq.~\eqref{eq:bound2}, but the bound $r^2 + 1$ from Proposition~\ref{thm:vc} already suffices to establish that $\ell$ is a tunable hyperparameter that controls the VC dimension.

\item For the third example, we use symmetry considerations as a tool to control the VC dimension.
First, partition the $n$-qubit register into disjoint subsets $I_1, \dots, I_k$ of size $|I_j| = m_j$ (i.e., $\sum_j m_j = n$).
Next, consider ``permutation-symmetry preserving'' parameterized unitaries on these partitions, which are defined as
\begin{align*}
    S^{+}_{I_j}(\theta) = e^{i\theta \sum_{i \in I_j}P_i},\quad \text{ and }\quad S^{\otimes}_{I_j}(\theta) = e^{i\theta \prod_{i \in I_j}P_i},
\end{align*}
where we have say $P_i = X_i$, $P_i = Y_i$, $P_i = Z_i$ or $P_i = I$ for all $i \in I_j$ (i.e., the same operator acting on all qubits in the partition $I_j$).
Note that if we apply these operators to a permutation invariant state on the $m_j$-qubits in the $j$th partition, then it remains permutation invariant (independent of $\theta$).
From these symmetric parameterized unitaries we construct parameterized layers $U(\theta_1, \dots, \theta_k) = \prod_{j=1}^k S_{I_j}^{+/\otimes}(\theta_j)$, from which we construct the ansatz as
\[
W(\theta) = U(\theta^d_1, \dots, \theta^d_k) \cdot \dots \cdot  U(\theta^1_1, \dots, \theta^1_k), \quad \theta \in [0, 1\pi)^{dk}.
\]
By the bound $r^2 + 1$ from Proposition~\ref{thm:vc}, for this choice of ansatz if the final measurement projects onto $\ell$ computational basis states the VC dimension is at most $r^2 + 1$, where 
\[
r = \ell \cdot \prod_{j=1}^k(m_j+1).
\]
This bound is obtained by computing the quantity in Eq.~\eqref{eq:bound1}, which can be done by noting that if we apply a layer $U$ to an $n$-qubit state that is invariant under permutations that only permute qubits within each partition, then it remains invariant under these permutations (i.e., independent of the choice of $\theta$).
In other words, the first column of $W(\theta)$ is always contained in the space of $n$-qubit states that are invariant under permutations that only permute qubits within each partition.
Next, note that the dimension of the space of $n$-qubit states that are invariant under permutations that only permute qubits within each partition is equal to $\prod_{j=1}^k (m_j + 1)$.
Finally, note that any other column of $W(\theta)$ spans a space whose dimension is at most that of the first column of $W(\theta)$ when varying $\theta$. 
Thus, any $\ell$ columns of $W(\theta)$ span a space of dimension is most $r = \ell \cdot \prod_{j=1}^k(m_j+1)$ when varying $\theta$.
Equivalent to the example above, for this particular choice of ansatze, computing the quantity in Eq.~\eqref{eq:bound2} might also be possible, but it is again a bit more involved and not necessary for our main goal of establishing that $\ell$ controls the VC dimension.
In particular, one might be able to compute the quantity in Eq.~\eqref{eq:bound2}, but the bound $r^2 + 1$ from Proposition~\ref{thm:vc} again already suffices to establish that $\ell$ is a tunable hyperparameter that controls the VC dimension.
\end{itemize}

In all of the above cases we see that we can control the upper bound on the VC dimension by varying the rank of the final measurement $\ell$.
It is worth noting that in these cases the regularized explicit quantum linear classifiers will generally give rise to a different model then the implicit approach without any theoretical guarantee regarding which will do better, because the standard relationship between the two models~\cite{schuld:kernel} will not hold anymore (i.e., the regularized explicit model does not necessarily correspond to a kernel method anymore).

Secondly, recall that by tuning the Frobenius norms of the observables used by a quantum linear classifier, we can balance the trade-off between its fat-shattering dimension and its ability to achieve large margins.
In particular, this shows that we can implement structural risk minimization of quantum linear classifiers with respect to the fat-shattering dimension by regularizing the Frobenius norms of the observables.
Again, it is important to note that the Frobenius norm itself does not fully characterize the generalization performance, since one also has to take into account the functional margin on training examples.
In particular, to optimize the generalization performance one has to minimize the Frobenius norm, while ensuring that the functional margin on training examples stays large.
As mentioned earlier, one way to achieve this is by maximizing the geometric margin, which on a set of examples $\{x_i\}$ is given by $\min_i \big|\tr{\mathcal{O}\rho_{\Phi}(x)} - d\big| / ||\mathcal{O}||_F$.
As before, for explicit quantum linear classifiers, we can estimate the Frobenius norm by sampling random computational basis states and computing the average of the postprocessing function $\lambda$ on them in order to estimate $||\mathcal{O}_{\theta}^\lambda||_F = \sqrt{\sum_{i = 1}^{2^n}\lambda(i)^2}$ (note that in some cases the Frobenius norm can be computed more directly).
On the other hand, for implicit quantum linear classifiers, we can regularize the Frobenius norm by regularizing $||\alpha||_1$ as $||\mathcal{O}_\alpha||_F \leq ||\alpha||_1$.
However, if the weights are obtained by solving the usual quadratic program~\cite{havlivcek:qsvm, schuld:qsvm}, then the resulting observable is already (optimally) regularized with respect to the Frobenius norm~\cite{schuld:kernel}.

Besides the types of regularization for which we have established theoretical evidence of the effect on structural risk minimization, there are also other types of regularization that are important to consider.
For instance, for explicit quantum linear classifiers, one could regularize the angles of the parameterized quantum circuit~\cite{park:practical}.
Theoretically analyzing the effect that regularizing the angles of the parameterized quantum circuit has on structural risk minimization would constitute an interesting direction for future research.
Another example is regularizing circuit parameters such as depth, width and number of gates for which certain theoretical results are known~\cite{bu:power1, caro:power}.
Finally, it turns out that one can also regularize quantum linear classifiers by running the circuits under varying levels of noise~\cite{bu:power3}.
For these kinds of regularization the relationships between the regularized explicit and regularized implicit quantum linear classifiers are still to be investigated.

\subsubsection*{Acknowledgments}
The results of this work extend on the MSc thesis of Dyon van Vreumingen~\cite{vreumingen:msc}.
The authors thank Matthias C. Caro, Maria Schuld and Ryan Sweke for giving valuable comments on the manuscript.
The authors thank Jordi Tura Brugu\'{e}s for discussions on the permutation-symmetry preserving ansatz.
This work was supported by the Dutch Research Council (NWO/OCW), as part of the Quantum Software Consortium programme (project number 024.003.037).

\bibliographystyle{quantum}
\bibliography{main}

\newpage 

\appendix

\section{Proofs of Section~\ref{subsec:fs-dim}}
\label{appendix:proofs_complexity}

\subsection{Proofs of Proposition~\ref{thm:vc} and Lemma~\ref{lemma:quadratic}
\label{appendix:vc}}

\vcdim*
\begin{proof}
Define $V = \sum_{\mathcal{O} \in \mathbb{O}}\mathrm{Im}\mathcal{O} \subset \C^{2^n}$ and let $P_V$ denote the orthogonal projector onto $V$.
Let $\Phi: \mathcal{X} \rightarrow \mathrm{Herm}\big(\C^{2^n}\big)$ denote the feature map of $\mathcal{C}_{\mathrm{qlin}}^{\mathbb{O}}$ and define $\Phi' = P_V\Phi P_V$.
Note that $\mathcal{C}_{\mathrm{qlin}}^{\mathbb{O}}(\Phi') = \mathcal{C}_{\mathrm{qlin}}^{\mathbb{O}}(\Phi)$. 
It is known that the VC dimension of linear classifiers on $\mathbb{R}^\ell$ is $\ell + 1$, and it is clear that $\mathrm{Herm}\big(V\big) \simeq \mathrm{Herm}\big(\C^r\big)\simeq\R^{r^2}$. 
Also, note that $\mathrm{Span}\big(\mathbb{O}\big)$ is a subspace of  $\mathrm{Herm}\big(V\big)$.
We therefore conclude that
\begin{align*}
\mathrm{VC}\big(\mathcal{C}_{\mathrm{qlin}}^{\mathbb{O}}(\Phi)\big) = \mathrm{VC}\big(\mathcal{C}_{\mathrm{qlin}}^{\mathbb{O}}(\Phi')\big) &\leq  \mathrm{VC}\big(\text{linear classifiers on } \mathrm{Span}\big(\mathbb{O}\big)\big) = \dim\big(\mathrm{Span}\big(\mathbb{O}\big)\big) + 1\\
&\leq \mathrm{VC}\big(\text{linear classifiers on } \mathrm{Herm}\big(V \big)\simeq\R^{r^2}\big) = r^2+1. 
\end{align*}
\end{proof}

\medskip

\quadratic*
\begin{proof}
First, we note that $V$ is contained in the space of Hermitian operators on $H$.
Since the dimension of the space of Hermitian operators on $H$ is equal to $\dim(H)^2$, this implies that
\[
    \dim(V) \leq \dim(H)^2.
\]
Next, we fix a basis of $H$ which we denote $\{\ket{\psi_k}\}_{k=1}^{\dim(H)}$, where we each $\ket{\psi_k}$ is of the form $\ket{\psi_{i}(\theta)}$ for some $i \in \{1, \dots, L\}$ and $\theta \in \mathbb{R}^m$.
To show that $\dim(V) \geq \dim(H)$, we will show that the operators $\{\ket{\psi_k}\bra{\psi_k}\}_{k=1}^{\dim(H)} \subset V$ are linearly independent.
We do so by contradiction, i.e., we assume they are not linearly independent and show that this leads to a contradiction.
That is, we assume that there exists a $k' \in \{1, \dots, \dim(H)\}$ and $\{\alpha_k\}_{k \neq k'}\subset \mathbb{R}$ such that
\[
    \ket{\psi_k'}\bra{\psi_k'} = \sum_{k \neq k'}\alpha_k\ket{\psi_k}\bra{\psi_k}.
\]
This implies that
\begin{align*}
    \ket{\psi_k'} &= \big(\ket{\psi_k'}\bra{\psi_k'} \big)\ket{\psi_k'}\\
    &= \Big(\sum_{k \neq k'}\alpha_k\ket{\psi_k}\bra{\psi_k}\Big)\ket{\psi_k'}\\
    &= \sum_{k \neq k'}(\alpha_k\cdot \braket{\psi_k\mid\psi_k'})\ket{\psi_k},
\end{align*}
which shows that $\{\ket{\psi_k}\}_{k=1}^{\dim(H)}$ are not linearly independent.
This clearly contradicts the assumption that $\{\ket{\psi_k}\}_{k=1}^{\dim(H)}$ is basis of $H$.
We therefore conclude that the operators $\{\ket{\psi_k}\bra{\psi_k}\}_{k=1}^{\dim(H)} \subset V$ are linearly independent, which shows that $\dim(V) \geq \dim(H)$.
\end{proof}

\subsection{Relationship between VC dimension bound and ranks of the observables
\label{appendix:rvsr}}

In this section we discuss one possible way to relate the quantity $r$ in Proposition~\ref{thm:vc} with the ranks of the observables by considering the overlaps of the images of the observables.
Specifically, consider a family of observables $\{\mathcal{O}_i\}_{i = 1}^n$, where each observable is of rank $R$\footnote{The results in this section hold more generally for families with varying ranks, though for simplicity (and to more closely relate it to Proposition~\ref{prop:expressivity_zeromargin}) we assume all observables have some fixed rank $R$ (from which it should be clear how to adapt it to the case where the observables can have different ranks).}.
Next, define the quantities
\begin{align}
    \label{eq:intersections}
    I_i = \dim\left(\mathrm{Im}\hspace{2pt}\mathcal{O}_i \cap \left[\mathrm{Im}\hspace{2pt}\mathcal{O}_{i+1} + \dots + \mathrm{Im}\hspace{2pt}\mathcal{O}_n \right] \right)
\end{align}
and
\begin{align}
    \label{eq:overlap}
    O_i = R - I_i.
\end{align}
Note that $O_i$ measures the extent to which the image of the observable $\mathcal{O}_i$ overlaps with the images of the observables $\mathcal{O}_{i+1}, \dots, \mathcal{O}_{n}$.
Specifically, $O_i$ is equal to zero if the images are fully overlapping, and it is equal to $R$ if there is no overlap at all.
Now Lemma~\ref{lemma:rvsr} below provides a way to relate the quantity $r$ in Proposition~\ref{thm:vc} with the ranks of the observables $R$ and the overlaps of the images~$O_i$.
Note that we consider the case where the family of observables is finite, whereas in the case of explicit quantum linear classifiers this family is infinite.
However, since all images live in a finite dimensional space, summing only finitely many images is already sufficient.
More precisely, for any family of $n$-qubit observables $\mathbb{O}$ (possibly infinitely large) there exists a $\mathbb{O}' \subseteq \mathbb{O}$ with $|\mathbb{O}'| \leq 2^n$ and
\[
\sum_{\mathcal{O}' \in \mathbb{O}'} \mathrm{Im}\hspace{2pt}\mathcal{O}' = \sum_{\mathcal{O} \in \mathbb{O}} \mathrm{Im}\hspace{2pt}\mathcal{O}.
\]
In Lemma~\ref{lemma:rvsr} below we can thus w.l.o.g.\ consider the case where the family of observables is finite.
\begin{lemma}
\label{lemma:rvsr}
Consider a family of observables $\mathbb{O} = \{\mathcal{O}_i\}_{i \in I}$, where each observable is of rank $R$.
Then, for $r$ defined in Proposition~\ref{thm:vc} and $\{O_i\}_{i \in I}$ defined in Eq.~\eqref{eq:overlap}, we have that
\[
r = R + \sum_{i=1}^{n-1}O_i
\]
\end{lemma}
\begin{proof}
The proof is basically a repeated application of the formula
\[
\dim\left(\mathrm{Im}\hspace{2pt}\mathcal{O}_1 + \mathrm{Im}\hspace{2pt}\mathcal{O}_2\right) = \dim\left(\mathrm{Im}\hspace{2pt}\mathcal{O}_1\right) + \dim\left(\mathrm{Im}\hspace{2pt}\mathcal{O}_2\right) - \dim\left(\mathrm{Im}\hspace{2pt}\mathcal{O}_1 \cap \mathrm{Im}\hspace{2pt}\mathcal{O}_2\right).
\]
Specifically, by repeatedly applying the above formula we find that
\begin{align*}
    r = \dim\left(\sum_{i = 1}^n \mathrm{Im}\hspace{2pt}\mathcal{O}_i\right) &= \dim\left(\mathrm{Im}\hspace{2pt}\mathcal{O}_1\right) + \dim\left(\sum_{i=2}^n\mathrm{Im}\hspace{2pt}\mathcal{O}_i\right) - \dim\left(\mathrm{Im}\hspace{2pt}\mathcal{O}_1 \cap \sum_{i=2}^n\mathrm{Im}\hspace{2pt}\mathcal{O}_i\right)\\
    &= \dim\left(\mathrm{Im}\hspace{2pt}\mathcal{O}_1\right) + \dim\left(\mathrm{Im}\hspace{2pt}\mathcal{O}_2\right) + \dim\left(\sum_{i=3}^n\mathrm{Im}\hspace{2pt}\mathcal{O}_i\right) \\
    &\hspace{20pt}- \dim\left(\mathrm{Im}\hspace{2pt}\mathcal{O}_1 \cap \sum_{i=2}^n\mathrm{Im}\hspace{2pt}\mathcal{O}_i\right)
    - \dim\left(\mathrm{Im}\hspace{2pt}\mathcal{O}_2 \cap \sum_{i=3}^n\mathrm{Im}\hspace{2pt}\mathcal{O}_i\right)\\
    &= nR - \left(I_1 + \dots + I_{n-1}\right)\\
    &= R - \sum_{i=1}^{n-1}(R - I_i) = R - \sum_{i =1}^{n-1}O_i
\end{align*}
\end{proof}

\subsection{Proof of Proposition~\ref{thm:fs_upperbound}
\label{appendix:fs}}
\fsdimupperbound*
\begin{proof}
Due to the close relationship to standard linear classifiers, we can utilize previously obtained results in that context.
In particular, for our approach we use the following proposition.
\begin{proposition}[Fat-shattering dimension of linear functions~\cite{shawe:generalization}]
\label{prop:fs_svm-1}
Consider the family of real-valued functions on the ball of radius $R$ inside $\mathbb{R}^N$ given by
\[
    \mathcal{F}_{\mathrm{lin}} = \Big\{ f_{w, d}(x)=\inp{w}{x} - d \text{ }\Big|\text{ }w\in \R^N\text{ with }|| w|| = 1,\text{ }d \in \mathbb{R}\text{ with }|d| \leq R \Big\}.
\]
The fat-shattering dimension of $\mathcal{F}_{\mathrm{lin}}$ can be bounded by
\[
    \fat_{\mathcal{F}_{\mathrm{lin}}}(\gamma)\leq \min\{9R^2/\gamma^2, N+1\}+1.
\]
\end{proposition}

The context in the above proposition is closely related, yet slightly different than that of quantum linear classifiers.
Firstly, $n$-qubit density matrices lie within the ball of radius $R=1$ inside $\mathrm{Herm}\big(\C^{2^n}\big)$ equipped with the Frobenius norm.
However, as in our case the hyperplanes arise from the family of observables $\mathbb{O}$, whose Frobenius norms are upper bounded by $\eta$, we cannot directly apply the above proposition.
We therefore adapt the above proposition by exchanging the role of $R$ with the upper bound on the norms of the observables in $\mathbb{O}$, resulting in the following lemma.
\begin{lemma}
\label{lemma:fs_svm-2}
Consider the family of real-valued functions on the ball of radius $R=1$ inside $\mathbb{R}^N$ given by
\[
    \mathcal{F}_{\mathrm{lin}}^{\leq \eta} = \Big\{ f_{w, d}(x)=\inp{w}{x} - d \text{ }\Big|\text{ }w\in \R^N\text{ with }||w|| \leq \eta ,\text{ }d \in \R\text{ with }|d| \leq \eta \Big\}.
\]
The fat shattering dimension of $\mathcal{F}_{\mathrm{lin}}^{\leq \eta}$ can be upper bounded by
\[
    \fat_{\mathcal{F}_{\mathrm{lin}}^{\leq \eta}}(\gamma)\leq\min\{9\eta^2/\gamma^2, N+1\}+1.
\]
\end{lemma}
\begin{proof}
Let us first determine the fat-shattering dimension of the family of linear functions with norm precisely equal to $\eta$ on points that lie within the ball of radius $R = 1$, i.e.,
\[
    \mathcal{F}_{\text{lin}}^{=\eta} = \Big\{ f_{w, d}(x)=\inp{w}{x} - d \text{ }\Big|\text{ }w\in \R^N\text{ with }||w|| = \eta ,\text{ }d \in \R\text{ with }|d| \leq \eta \Big\}.
\]
Suppose $\mathcal{F}_{\text{lin}}^{=\eta}$ can $\gamma$-shatter a set of points $\{x_1, \dots, x_k\}$ that lie within the ball of radius $R=1$.
Because $\inp{w}{x_i} = \inp{w / \eta}{ \eta x_i}$, we find that $\mathcal{F}_{\text{lin}}^{=1}$ can $\gamma$-shatter the set of points $\eta x_1, \dots, \eta x_k$ that lie within the ball of radius $R = \eta$.
By Proposition~\ref{prop:fs_svm-1} we have $k \leq \min\{9\eta^2/\gamma^2, N+1\}+1$.
Thus, the fat-shattering dimension of $\mathcal{F}_{\text{lin}}^{=\eta}$ on points within the ball of radius $R=1$ is upper bounded by
\[
    \fat_{\mathcal{F}_{\text{lin}}^{=\eta}}(\gamma) \leq \min\{9\eta^2/\gamma^2, N+1\}+1.
\]
To conclude the desired results, note that this bound is monotonically increasing in $\eta$, and thus allowing hyperplanes with with norm $\|w\|<\eta$ will not increase the fat-shattering dimension.

\end{proof}
From the above lemma we can immediately infer an upper bound on the fat-shattering dimension of quantum linear classifiers by identifying that as vector spaces $\mathrm{Herm}\big(\C^{2^n}\big) \simeq \R^{4^n}$.

\end{proof}

\subsubsection{Sample complexity in the PAC-learning framework}

Besides being related to generalization performance, the fat-shattering dimension is also related to the so-called \emph{sample complexity} in the probably approximately correct (PAC) learning framework~\cite{kearns:fs}. 
The sample complexity captures the amount classifier queries required to find another classifier that with high probability agrees with the former classifier on unseen examples.

By plugging the upper bound of Proposition~\ref{thm:fs_upperbound} into previously established theorems on the sample complexity of families of classifiers~\cite{anthony:pac, bartlett:pac}, we derive the following corollary, which can be viewed as a dual of the result of~\cite{aaronson:learnqstates}.
\begin{corollary}
\label{cor:pac}
Let $\mathbb{O}\subseteq \mathrm{Herm}\left(\mathbb{C}^{2^n} \right)$ be a family of observables with $\eta = \max_{\mathcal{O}\in \mathbb{O}} \|\mathcal{O}\|_F$ and consider the family of real-valued functions $\mathcal{F}_{\mathrm{qlin}}^{\mathbb{O}}$ defined in Eq.~\eqref{eq:f_qlin_O}. 
Fix an element $F \in \mathcal{F}_{\mathrm{qlin}}^{\mathbb{O}}$ as well as parameters $\epsilon, \nu, \gamma > 0$ with $\gamma\epsilon \geq 7 \nu$. 
Suppose we draw $m$ examples $\mathcal{D} = \{\rho_1, \dots, \rho_m\}$ independently according to a distribution $P$, and then choose any function $H \in \mathcal{F}_{\mathrm{qlin}}^{\mathbb{O}}$ such that $| H(\rho_i) - F(\rho_i)| \leq \nu$ for all $\rho_i \in \mathcal{D}$. 
Then, with probability at least $1-\delta$ over $P$, we have that
\begin{align*}
\underset{\rho \sim P}{\mathrm{Pr}}\big(|H(\rho) - F(\rho)| > \gamma \big) \leq \epsilon,
\end{align*}
provided that
\begin{align*}
m \in \Omega\bigg(\frac{1}{\gamma^2 \epsilon^2}\Big(\frac{\eta^2}{\gamma^2\epsilon^2}\log^2\frac{1}{\gamma\epsilon} + \log\frac{1}{\delta}\Big)\bigg).
\end{align*}
\end{corollary}

\begin{proof}
Follows directly from plugging the uppper bound of Proposition~\ref{thm:fs_upperbound} into Corollary 2.4 of~\cite{aaronson:learnqstates}.

\end{proof}

\section{Proofs of propositions Section~\ref{subsec:expressivity}}
\label{appendix:proofs_empirical}

\subsection{Proof of Proposition~\ref{prop:expressivity_zeromargin}
\label{appendix:rank}}
\zeromargin*
\begin{proof}
\emph{(i):} Suppose $c_{\mathcal{O}, b} \in \mathcal{C}_{\mathrm{qlin}}^{(k)}$ correctly classifies $\mathcal{D}$.
Let $\delta = \min_{x \in \mathcal{D}_{-}}\big|\tr{\mathcal{O}\rho_x} - d \big|$, where $\mathcal{D}_{-}$ is the subset of examples with label $-1$, and note that since $\mathcal{D}$ is correctly classified we have $\delta > 0$.
Fix the basis we work in to be the eigenbasis of $\mathcal{O}$ ordered in such a way that 
\[
\mathcal{O} = \mathrm{diag}(\lambda_1, \dots, \lambda_k, 0, \dots, 0)
\]
and define
\[
P = \frac{1}{r-k}\mathrm{diag}(\underbrace{0, \dots, 0}_{k \text{ times}}, \underbrace{1, \dots, 1}_{r-k \text{ times}}, \underbrace{0, \dots, 0}_{2^n - r \text{ times}}).
\]
For every $0 < \epsilon < \delta$ we have that $\mathcal{O}' = \mathcal{O} + \epsilon P$ has $\mathrm{rank}(\mathcal{O}') = r$.
What remains to be shown is that $c_{\mathcal{O}', b} \in \mathcal{C}^{(r)}_{\mathrm{qlin}}$ correctly classifies $\mathcal{D}$.
To do so, first let $x \in \mathcal{D}_+$ (i.e., labeled $+1$) and note that
\[
    \tr{\mathcal{O}'\rho_{x}} - b = \underbrace{(\tr{\mathcal{O}\rho_{x}} - b)}_{\geq 0} + \underbrace{\epsilon\tr{P\rho_{x}}}_{\geq 0} \geq 0,
\]
which shows that indeed $c_{\mathcal{O}', b}(x) = +1$.
Next, let $x \in \mathcal{D}_-$ (i.e., labeled $-1$) and note that
\[
    \tr{\mathcal{O}'\rho_{x}} - b = \underbrace{(\tr{\mathcal{O}\rho_{x}} - b)}_{\leq -\delta} + \underbrace{\epsilon\tr{P\rho_{x^+}}}_{< \delta} < 0,
\]
which shows that indeed $c_{\mathcal{O}', b}(x) = -1$.

\medskip

\emph{(ii):}

We will describe a protocol that queries a classifier $c_{\mathcal{O}, b}$ and based on its outcomes checks whether $\mathcal{O}$ is approximately equal to a fixed target observable $\mathcal{T}$ of rank $r$.
We will show that if the queries to $c_{\mathcal{O}, b}$ are labeled in a way that agrees with the target classifier that uses the observable $\mathcal{T}$, then the spectrum of $\mathcal{O}$ has to be point-wise within distance $\epsilon$ of the spectrum of $\mathcal{T}$.
In particular, this will show that the rank of $\mathcal{O}$ has to be at least $r$ if we make $\epsilon$ small enough.
Consequently, if the rank of $\mathcal{O}$ is less than $r$, then at least one query made during the protocol has to be labeled differently by $c_{\mathcal{O}, b}$ than the target classifier.
In the end, the queries made to the classifier during the protocol will therefore constitute the set of examples described in the theorem.

Let us start with some definition. 
For a classifier $c_{\mathcal{O}, b}(\rho) = \text{sgn}\big(\text{Tr}\big[\mathcal{O}\rho\big] - b\big)$ we define its effective observable $\mathcal{O}_{\text{eff}} = \mathcal{O} - bI$ which we express in the computational basis as $\mathcal{O}_{\text{eff}} = (O_{ij})$.
Next, we define our target classifier to be $c_{\mathcal{T}, -1}$ where the observable $\mathcal{T}$ is given by
\[
    \mathcal{T} = -r \cdot \ket{0}\bra{0} + \sum_{i = 1}^{r-1} i \cdot \ket{i}\bra{i},
\]
and we define its effective observable $\mathcal{T}_{\text{eff}} = \mathcal{T} + I$ which we express in the computational basis as $\mathcal{T}_{\text{eff}} = (T_{ij})$.
Rescaling $\mathcal{O}_{\mathrm{eff}}$ with a positive scalar does not change the output of the corresponding classifier. 
Therefore, to make the protocol well-defined, we define $\mathcal{O}_{\mathrm{eff}}$ to be the unique effective observable whose first diagonal element is scaled to be equal to $O_{00} = -(r+1)$. 

Our approach is as follows.
First, we query $c_{\mathcal{O}, b}$ in such a way that if the outcomes agree with with the target classifier $c_{\mathcal{T}, -1}$, then the absolute values of the off-diagonal entries in the first row and column of $\mathcal{O}_{\text{eff}}$ must be close to zero (i.e., approximately equal to those of $\mathcal{T}_{\text{eff}}$).
Afterwards, we again query $c_{\mathcal{O}, b}$ but now in such a way that if the outcomes agree with the target classifier $c_{\mathcal{T}, -1}$, then the diagonal elements of $\mathcal{O}_{\text{eff}}$ must be approximately equal to those of $\mathcal{T}_{\text{eff}}$.
In the end, we query $c_{\mathcal{O}, b}$ one final time but this time in such a way that if the outcomes agree with the target classifier $c_{\mathcal{T}, -1}$, then the absolute values of the remaining off-diagonal elements of $\mathcal{O}_{\text{eff}}$ must be close to zero (i.e., again approximately equal to those of $\mathcal{T}_{\text{eff}}$).
Finally, we use Gershgorin's circle theorem to show that the spectrum of $\mathcal{O}_{\text{eff}}$ has to be point-wise close to the spectrum of $\mathcal{T}_{\text{eff}}$.
We remark that this procedure could be generalized to a more complete tomography approach, where one uses queries to the classifier $c_{\mathcal{O}, b}$ in order to reconstruct the entire spectrum of $\mathcal{O}_{\text{eff}}$.

First, we query the quantum states $\ket{i}$ for $i = 0, \dots, 2^n-1$. 
Without loss of generality, we can assume that the classifiers $c_{\mathcal{O}, b}$ and $c_{\mathcal{T}, -1}$ agree on the label, i.e., 
\begin{align}
    \label{eq:diag_signs}
    c_{\mathcal{O}, b}\big(\ket{0}\bra{0}\big) = -1,\text{ and }c_{\mathcal{O}, b}\big(\ket{i}\bra{i}\big) = +1 \enskip\text{for $i=1, \dots, 2^n-1$},
\end{align}
as otherwise a set of examples containing just these states would already separate $c_{\mathcal{O}, b}$ and $c_{\mathcal{T}, -1}$.

In order to show that the absolute value of the off-diagonal elements of the first row and column of $\mathcal{O}_{\text{eff}}$ must be close to zero and that the diagonal elements of $\mathcal{O}_{\text{eff}}$ must be close to those of $\mathcal{T}_{\text{eff}}$, we consider the quantum states given by
\begin{align}
\label{eq:qstate_gamma}
\ket{\gamma_\theta (\alpha)} = \sqrt{1- \alpha}\ket{0} + e^{i\theta}\sqrt{\alpha}\ket{j}, \quad \text{ with }\alpha \in [0, 1]\text{ and }\theta \in [0, 2\pi).
\end{align}
Its expectation value with respect to $\mathcal{O}_{\text{eff}}$ is given by
\begin{align}
\label{eq:expectation_gamma_O}
\bra{\gamma_\theta (\alpha)}\mathcal{O}_{\text{eff}}\ket{\gamma_\theta (\alpha)} = (1-\alpha)\cdot O_{00} + \alpha\cdot O_{jj} + \sqrt{\alpha(1- \alpha)}\cdot C_{\theta}, \quad\text{ where }C_{\theta} := \text{Re}\big(e^{i\theta} O_{0j}\big),
\end{align}
and its expectation value with respect to $\mathcal{T}_{\text{eff}}$ is given by
\begin{align}
\label{eq:expectation_gamma_T}
\bra{\gamma_\theta (\alpha)}\mathcal{T}_{\text{eff}}\ket{\gamma_\theta (\alpha)} &= (1-\alpha)\cdot T_{00} + \alpha\cdot T_{jj}.
\end{align}
Crucially, by Equation~\eqref{eq:diag_signs} we know that the label of $\ket{\gamma_\theta (\alpha)}$ goes from $-1$ to $+1$ as $\alpha$ goes $0 \rightarrow 1$.
Note that the expectation value of $\ket{\gamma_\theta (\alpha)}$ with respect to $\mathcal{T}_{\text{eff}}$ is independent from the phase $\theta$.

To determine that $\big|O_{0j}\big|$ is smaller than $\delta >0$, we query the classifier $c_{\mathcal{O},b}$ on the states $\ket{\gamma_{\hat{\theta}}(\hat{\alpha})}$ for all $\hat{\theta}$ in a $\zeta$-mesh of $[0, 2\pi)$ and for all $\hat{\alpha}$ in a $\xi$-mesh of $[0,1]$ and we suppose they are labeled the same as the target classifier $c_{\mathcal{T},-1}$ would label them.
Using these queries we can find estimates $\hat{\alpha}^{\mathcal{O}_{\text{eff}}}_{\text{cross}}(\hat{\theta})$ that are $\xi$-close to the unique $\alpha^{\mathcal{O}_{\text{eff}}}_{\text{cross}}(\theta) = \alpha'$ that satisfies
\begin{align}
\label{eq:crossing_alpha}
\bra{\gamma_\theta (\alpha')}\mathcal{O}_{\text{eff}}\ket{\gamma_\theta (\alpha')} = 0,
\end{align}
by finding the smallest $\hat{\alpha}$ where the label has gone from $-1$ to $+1$. 
We refer to the $\alpha'$ satisfying Equation~\eqref{eq:crossing_alpha} as the \emph{crossing point at phase $\theta$}.
Because the label assigned by $c_{\mathcal{T}, -1}$ does not depend on the phase $\theta$, and since all states $\ket{\gamma_{\hat{\theta}}(\hat{\alpha})}$ were assigned the same label by $c_{\mathcal{O}, b}$ and $c_{\mathcal{T}, -1}$, we find that the crossing point estimate $\hat{\alpha}^{\mathcal{O}_{\text{eff}}}_{\text{cross}}(\hat{\theta})$ is the same for all $\hat{\theta}$.
In particular, this implies that the actual crossing points $\alpha^{\mathcal{O}_{\text{eff}}}_{\text{cross}}(\hat{\theta})$ have to be within $\xi$-distance of each other for all $\hat{\theta}$.

Before we continue, we first show that if $c_{\mathcal{O}, b}$ assigns the same labels as $c_{\mathcal{T}, -1}$, then $O_{jj}$ is bounded above by a quantity that only depends on $n$.
Fix $\tilde{\theta}$ to be any point inside the $\zeta$-mesh such that $C_{\tilde{\theta}} \leq 0$, and define the function $E(\alpha) = (1-\alpha)\cdot O_{00} + \alpha\cdot O_{jj} + \sqrt{(1-\alpha)\alpha}\cdot C_{\tilde{\theta}}$.
By our choice of $\mathcal{T}$, we have that $\alpha_{\text{cross}}^\mathcal{T} \in (\frac{r+1}{2r+1}, \frac{r+1}{r+3})$. Therefore, if $c_{\mathcal{O}, b}$ and $c_{\mathcal{T}, -1}$ agree on the entire $\xi$-mesh for a small enough $\xi$, then it must hold that $\alpha^{\mathcal{O}_{\text{eff}}}_{\text{cross}}(\tilde{\theta}) \in (\frac{1}{2}, \frac{2^n+1}{2^n+2})$.
By the mean value theorem there exists an $\alpha' \in (\alpha^{\mathcal{O}_{\text{eff}}}_{\text{cross}}(\tilde{\theta}), \frac{2^n+1}{2^n+2})$ such that 
\begin{align}
\label{eq:boundjjintermediate}
E'(\alpha') = \frac{E(\frac{2^n+1}{2^n+2}) - E(\alpha^{\mathcal{O}_{\text{eff}}}_{\text{cross}}(\tilde{\theta}))}{\frac{2^n+1}{2^n+2} - \alpha^{\mathcal{O}_{\text{eff}}}_{\text{cross}}(\tilde{\theta})}.
\end{align}
After some rewriting, we can indeed conclude from the above equation that $O_{jj}$ is bounded above by a quantity that only depends on $n$.

Next, write $O_{0j} = \big|O_{0j}\big|e^{i \phi}$ with $\phi \in [0, 2\pi)$, let $\hat{\theta}_{\text{abs}}$ denote the point in the $\zeta$-mesh of $[0,2\pi)$ that is closest to $2\pi - \phi$, and let $\hat{\theta}_{0}$ denote the point in the $\zeta$-mesh of $[0,2\pi)$ that is closest to $\pi/2 - \phi$ modulo $2\pi$.
By our previous discussion we know that $\big|\alpha^{\mathcal{O}_{\text{eff}}}_{\text{cross}}(\hat{\theta}_{\text{abs}}) - \alpha^{\mathcal{O}_{\text{eff}}}_{\text{cross}}(\hat{\theta}_{0}) \big| < \xi$, 
which together with the previously established bound on $O_{jj}$ implies that
\begin{align}  
\label{eq:cross_upper}
    \Big|C_{\hat{\theta}_{\text{abs}}} - C_{\hat{\theta}_{0}}\Big| < f(\xi),
\end{align}
where $f$ is a continuous function (independent from $c_{\mathcal{O}, b}$) with $f(\xi) \rightarrow 0$ as $\xi \rightarrow 0$.   
Moreover, using the inequality $\cos(\zeta) \geq 1 - \lambda\zeta$, where $\lambda\approx 0.7246$ is a solution of $\lambda\big(\pi - \arcsin(\lambda)\big) = 1 + \sqrt{1 - \lambda^2}$, together with the inequality $\cos(\pi/2 - \zeta) \leq \zeta$, we can derive that
\begin{align}
\label{eq:cross_lower}
\begin{split}
\Big|C_{\hat{\theta}_{\text{abs}}} - C_{\hat{\theta}_{0}}\Big| &= \Big|\big|O_{0j}\big|\cos\big(\hat{\theta}_{\text{abs}} + \phi\big) - \big|O_{0j}\big|\cos\big(\hat{\theta}_{0} + \phi\big)\Big| \\
&\geq \Big|O_{0j}\Big|\cdot \Big|\cos\big(\zeta\big) - \cos\big(\pi/2 - \zeta\big) \Big| \\
&\geq \Big|O_{0j}\Big|\cdot \Big| 1 - \big(\lambda+1\big)\zeta \Big|.
\end{split}
\end{align}
Finally, by combining Equation~\eqref{eq:cross_upper} with Equation~\eqref{eq:cross_lower} we can conclude that
\[
    \big|O_{0j}\big| < \frac{f(\xi)}{1 - (\lambda + 1) \zeta},
\]
which for $\xi$ and $\zeta$ small enough shows that $\big|O_{0j}\big| < \delta$ for any chosen precision $\delta > 0$ (i.e., the fineness of both meshes $\xi$ and $\zeta$ will depend on the choice of $\delta$).

To determine that $O_{jj}$ is within distance $\delta'>0$ of $T_{jj}$ we again query the classifier $c_{\mathcal{O},b}$ but this time on the states $\ket{\gamma_{0}(\hat{\alpha})}$ for all $\hat{\alpha}$ in a $\xi'$-mesh of $[0,1]$ and we suppose they are labeled the same as the target classifier $c_{\mathcal{T},-1}$ would.
Using these queries we can find estimates $\hat{\alpha}^{\mathcal{O}_{\text{eff}}}_{\text{cross}}(0)$, $\hat{\alpha}^{\mathcal{T}_{\text{eff}}}_{\text{cross}}(0)$ that are $\xi'$-close to the corresponding actual crossing point.
As we assumed that all queries are labeled the same by $c_{\mathcal{O}, b}$ and $c_{\mathcal{T}, -1}$, the crossing point estimate $\hat{\alpha}^{\mathcal{O}_{\text{eff}}}_{\text{cross}}(0)$ has to be equal to the crossing point estimate $\hat{\alpha}^{\mathcal{T}_{\text{eff}}}_{\text{cross}}(0)$.
In particular, this implies that the actual crossing points $\alpha^{\mathcal{O}_{\text{eff}}}_{\text{cross}}(0)$ and  $\alpha^{\mathcal{T}_{\text{eff}}}_{\text{cross}}(0)$ have to be within $\xi'$-distance of each other. 
Next, define $g(\alpha, C)$ to be the unique coefficient $O \in \mathbb{R}_{\geq 0}$ that satisfies
\[
   (1-\alpha)\cdot O_{00} + \alpha\cdot O + \sqrt{\alpha(1-\alpha)}\cdot C = 0.
\]
It is clear that $g$ is a continuous function in $\alpha$ and $C$ that is independent from $c_{\mathcal{O}, b}$, and that $T_{jj} = g(\alpha^{\mathcal{T}_{\text{eff}}}_{\text{cross}}(0), 0)$ and $O_{jj} = g(\alpha^{\mathcal{O}_{\text{eff}}}_{\text{cross}}(0), C_{0})$.
Finally, we let $\delta >0$ and $\xi'> 0$ be small enough such that if $\big|\alpha^{\mathcal{O}_{\text{eff}}}_{\text{cross}}(0) -\alpha^{\mathcal{T}_{\text{eff}}}_{\text{cross}}(0)\big| < \xi'$ and $\big|C_0\big| < \delta$, then 
\[
    \big|O_{jj} - T_{jj} \big| = \big| g(\alpha^{\mathcal{T}_{\text{eff}}}_{\text{cross}}(0), 0) - g(\alpha^{\mathcal{O}_{\text{eff}}}_{\text{cross}}(0), C_{0}) \big| < \delta'.
\]
In conclusion, to determine that $O_{jj}$ is within distance $\delta'>0$ of $T_{jj}$ we first do the required queries to determine that $\big|C_0\big| = \big|O_{0j}\big|< \delta$, after which we do the required queries to determine that $\big|\alpha^{\mathcal{O}_{\text{eff}}}_{\text{cross}}(0) -\alpha^{\mathcal{T}_{\text{eff}}}_{\text{cross}}(0)\big| < \xi'$, which together indeed implies that $O_{jj}$ is within distance $\delta'>0$ of $T_{jj}$.   

In order to show that the absolute value of the remaining off-diagonal elements of $\mathcal{O}_{\text{eff}}$ must be close to zero (i.e., close to those of $\mathcal{T}_{\text{eff}}$) we consider the quantum states given by
\begin{align}
\label{eq:qstate_mu}
\ket{\mu_\theta(\alpha)} = \frac{\sqrt{1-\alpha}}{\sqrt{2}}\big(\ket{0} + \ket{i}\big) + e^{i\theta}\sqrt{\alpha}\ket{j}, \quad \text{ with }\alpha \in [0, 1]\text{ and }\theta \in [0, 2\pi).
\end{align}
Its expectation value with respect to $\mathcal{O}_{\text{eff}}$ is given by
\begin{align}
\label{eq:expectation_mu_O}
 \bra{\mu_\theta(\alpha)}\mathcal{O}_{\text{eff}}\ket{\mu_\theta(\alpha)} =\big(1 - \alpha\big)\cdot\big(O_{00} + O_{ii} + \text{Re}(O_{0i})\big) + \alpha\cdot O_{jj} + \sqrt{2\alpha(1-\alpha)}\cdot C_{\theta},
\end{align}
where $C_{\theta} := \text{Re}\big(e^{i\theta} (O_{0j} + O_{ij})\big)$, and its expectation value with respect to $\mathcal{T}_{\text{eff}}$ is given by
\begin{align}
\label{eq:expectation_mu_T}
\bra{\mu_\theta(\alpha)}\mathcal{T}_{\text{eff}}\ket{\mu_\theta(\alpha)} =\big(1 - \alpha\big)\cdot\big(T_{00} + T_{ii}\big) + \alpha\cdot T_{jj}.
\end{align}
Crucially, by our choice of $\mathcal{T}$ we know that the label of $\ket{\mu_\theta (\alpha)}$ goes from $-1$ to $+1$ as $\alpha$ goes $0 \rightarrow 1$.
Note that the expectation value of $\ket{\mu_\theta (\alpha)}$ with respect to $\mathcal{T}_{\text{eff}}$ is independent from the phase $\theta$.

To determine that $\big|O_{ij}\big|$ is smaller than $\delta'' >0 $ for $i, j \geq 1$ and $i \neq j$, we query the classifier $c_{\mathcal{O},b}$ on the states $\ket{\gamma_{\hat{\theta}}(\hat{\alpha})}$ for all $\hat{\theta}$ in a $\zeta''$-mesh of $[0, 2\pi)$ and for all $\hat{\alpha}$ in a $\xi''$-mesh of $[0,1]$ and we suppose they are labeled the same as the target classifier $c_{\mathcal{T},-1}$ would.
Using these queries we can find estimates $\hat{\alpha}^{\mathcal{O}_{\text{eff}}}_{\text{cross}}(\hat{\theta})$ that are $\xi$-close to the unique $\alpha^{\mathcal{O}_{\text{eff}}}_{\text{cross}}(\theta) = \alpha'$ that satisfies
\begin{align}
\label{eq:crossing_mu}
\bra{\mu_\theta (\alpha')}\mathcal{O}_{\text{eff}}\ket{\mu_\theta (\alpha')} = 0,
\end{align}
by finding the smallest $\hat{\alpha}$ where the label has gone from $-1$ to $+1$. 
Because the label assigned by $c_{\mathcal{T}, -1}$ does not depend on the phase $\theta$, and since all states $\ket{\mu_{\hat{\theta}}(\hat{\alpha})}$ were assigned the same label by $c_{\mathcal{O}, b}$ and $c_{\mathcal{T}, -1}$, we find that the crossing point estimate $\hat{\alpha}^{\mathcal{O}_{\text{eff}}}_{\text{cross}}(\hat{\theta})$ is the same for all $\hat{\theta}$.
In particular, this implies that the actual crossing points $\alpha^{\mathcal{O}_{\text{eff}}}_{\text{cross}}(\hat{\theta})$ have to be within $\xi''$-distance of each other for all $\hat{\theta}$.
Subsequently, write $O_{0j} + O_{ij} = \big|O_{0j} + O_{ij}\big|e^{i \phi}$ with $\phi \in [0, 2\pi)$, let $\hat{\theta}_{\text{abs}}$ denote the point in the $\zeta''$-mesh of $[0,2\pi)$ that is closest to $2\pi - \phi$, and let $\hat{\theta}_{0}$ denote the point in the $\zeta''$-mesh of $[0,2\pi)$ that is closest to $\pi/2 - \phi$ modulo $2\pi$.
By our previous discussion we know that $\big|\alpha^{\mathcal{O}_{\text{eff}}}_{\text{cross}}(\hat{\theta}_{\text{abs}}) - \alpha^{\mathcal{O}_{\text{eff}}}_{\text{cross}}(\hat{\theta}_{0}) \big| < \xi''$, 
which implies
\begin{align}  
\label{eq:cross_upper_mu}
    \Big|C_{\hat{\theta}_{\text{abs}}} - C_{\hat{\theta}_{0}}\Big| < h(\xi''),
\end{align}
where $h$ is a continuous function (independent from $c_{\mathcal{O}, b}$ and $c_{\mathcal{T}, -1}$) with $h(\xi'') \rightarrow 0$ as $\xi'' \rightarrow 0$.   
Moreover, using the inequality $\cos(\zeta'') \geq 1 - \lambda\zeta''$, where $\lambda\approx 0.7246$ is a solution of $\lambda\big(\pi - \arcsin(\lambda)\big) = 1 + \sqrt{1 - \lambda^2}$, together with the inequality $\cos(\pi/2 - \zeta'') \leq \zeta''$, we can derive that
\begin{align}
\label{eq:cross_lower_mu}
\begin{split}
\Big|C_{\hat{\theta}_{\text{abs}}} - C_{\hat{\theta}_{0}}\Big| &= \Big|\big|O_{0j} + O_{ij}\big|\cos\big(\hat{\theta}_{\text{abs}} + \phi\big) - \big|O_{0j} + O_{ij}\big|\cos\big(\hat{\theta}_{0} + \phi\big)\Big| \\
&\geq \Big|O_{0j} + O_{ij}\Big|\cdot \Big|\cos\big(\zeta''\big) - \cos\big(\pi/2 - \zeta''\big) \Big| \\
&\geq \Big|O_{0j} + O_{ij}\Big|\cdot \Big| 1 - \big(\lambda+1\big)\zeta'' \Big|.
\end{split}
\end{align}
Finally, by combining Equation~\eqref{eq:cross_upper_mu} with Equation~\eqref{eq:cross_lower_mu} we can conclude that
\[
    \big|O_{0j} + O_{ij}\big| < \frac{h(\xi'')}{1 - (\lambda + 1) \zeta''},
\]
which for $\xi''$ and $\zeta''$ small enough shows that $\big|O_{0j} + O_{ij}\big| < \delta''/2$ (i.e., the fineness of both meshes $\xi''$ and $\zeta''$ will depend on the choice of $\delta''$).
In conclusion, to determine that $\big|O_{ij}\big|$ is smaller than $\delta''>0$ we first do the required queries to determine that $\big|O_{0j}\big| < \delta''/2$, after which we do the required queries to determine that $\big|O_{0j} + O_{ij}\big| < \delta''/2$, which together indeed implies that $\big|O_{ij}\big| < \delta''$.

All in all, we have described a (finite) set of states such that if the label assigned by $c_{\mathcal{O}, b}$ agrees with the label assigned by $c_{\mathcal{T}, -1}$, then the absolute value of the off-diagonal elements of the first row of $\mathcal{O}_{\text{eff}}$ have to be smaller than $\delta$, the diagonal elements of $\mathcal{O}_{\text{eff}}$ have to be within $\delta'$-distance of those of $\mathcal{T}_{\text{eff}}$, and the remaining off diagonal elements of $\mathcal{O}_{\text{eff}}$ have to be smaller than $\delta''$.
Finally, we choose $\delta,\delta',\delta'' = 1 / 2^{n+1}$ and use the above protocol to establish that for $1 \leq i \leq r-1$ the Gershgorin discs $D_i$ of $\mathcal{O}_{\text{eff}}$ (i.e., with center $O_{ii}$ and radius $\sum_j |O_{ij}|$) have to be contained in the disks $\tilde{D}_i$ with center $i+1$ and radius $1/2$.
Moreover, we establish that the Gershgorin disc $D_0$ has to be contained in the disks $\tilde{D}_0$ with center $-r + 1$ and radius $1/2$.
Since the disks $\tilde{D}_i$ as disjoint, so are the Gershgorin discs $D_i$, which implies that $\mathcal{O}_{\text{eff}}$ must have at least $r$ distinct eigenvalues, and thus that $\text{rank}\big( \mathcal{O}\big) \geq r$.
Consequently, if $\text{rank}\big( \mathcal{O}\big) < r$, then $c_{\mathcal{O}, b}$ must disagree with $c_{\mathcal{T}, -1}$ on the label of at least one of the states queried during the protocol.

\end{proof}

\subsection{Proof of Proposition~\ref{prop:comparison}
\label{appendix:comparisson}}
\comparison*
\begin{proof}
\emph{(i):}
First, we define the feature map $\Phi':\R^\ell \rightarrow \R^{N+1}$ which maps
\[
    x \mapsto \frac{\Phi(x)}{M} + \sqrt{1 - \frac{||\Phi(x)||^2}{M^2}}e_{N+1},
\]
where $e_{N+1}$ denotes the $(N+1)$-th standard basis vector.
Note that this feature map indeed satisfies that $||\Phi'(x)|| = 1$ for all $x \in \R^\ell$.
Next, for any classifier $c_{w, b} \in \mathcal{C}_{\mathrm{qlin}}(\Phi)$ we define $w' = w$ and $b' = b / M$ and we note that for any $x \in \R^\ell$ we have
\[
    c_{w', b'}(\Phi'(x)) = \sign{\inp{w'}{\Phi'(x)} - b'} = \sign{M^{-1}\big[\inp{w}{\Phi(x)} - b\big]} = \sign{\inp{w}{\Phi(x)} - b} = c_{w, b}(\Phi(x)).
\]

\emph{(ii):}
First, we define the feature map $\widetilde{\Phi}: \R^\ell \rightarrow \R^{N+1}$ which maps 
\[
    x \mapsto \Phi(x) + e_{N+1},
\]
where $e_{N+1}$ denotes the $(N+1)$-th standard basis vector.
Next, for any classifier $c_{w, b} \in \mathcal{C}_{\mathrm{lin}}(\Phi)$ we define $\tilde{w} = w - be_{N+1}$ and we note that for all $x \in \R^\ell$ we have
\[
    c_{\tilde{w}, 0}(\widetilde{\Phi}(x)) = \sign{\inp{\widetilde{\Phi}(x)}{\widetilde{w}}} = \sign{\inp{\Phi(x)}{w} - b} = c_{w, b}(\Phi(x)).
\]

Therefore, it suffices to show that we can implement any linear classifier on $\mathbb{R}^{N+1}$ with $b=0$ as a quantum linear classifier on $n = \lceil\log N+1\rceil + 1$ qubits.
To do so, we define the quantum feature map $\Phi' : \R^\ell \rightarrow \mathrm{Herm}\big(\C^{2^n}\big)$ which maps
\[
    x \rightarrow \rho_x = \left(\frac{\ket{\Phi(x)} + \ket{0}}{\sqrt{2}}\right)\left(\frac{\bra{\Phi(x)} + \bra{0}}{\sqrt{2}}\right),
\]
where $\ket{0}$ is a vector that does not lie in the support of $\Phi$ (note this vectors exists since we have chosen $n$ large enough).
Finally, for any linear classifier $c_{w, 0} \in \mathcal{C}_{\mathrm{lin}}(\Phi)$ on $\R^{N+1}$ we define $b' = ||w||^2/2$ and $\mathcal{O} = \ket{w'}\bra{w'}$, where $\ket{w'} = \ket{w} + ||w||\ket{0}$ and we note that for all $x \in \R$ we have
\begin{align*}
    c_{\mathcal{O}, b'}(\Phi'(x)) &= \sign{\tr{\mathcal{O}\rho_x} - b'} \\&
    = \sign{\frac{1}{2}\Big|\braket{w\mid\Phi(x)} + ||w||\text{ }\Big|^2 - \frac{||w||^2}{2}}\\
        &= \sign{\inp{w}{\Phi(x)}} = c_{w, 0}(\Phi(x)).
\end{align*}
\emph{(iii):} This follows directly from the fact that $\mathrm{Herm}\left( \C^{2^n}\right) \simeq \R^{4^n}$.

\end{proof}

\subsection{Proof of Proposition~\ref{prop:expressivity_nonzeromargin}
\label{appendix:fs_expr}}
\nonzeromargin*
\begin{proof}
Define $\mathcal{D}_m = \mathcal{D}_m^+ \cup \mathcal{D}_m^-$ whose positive examples (i.e., labeled $+1$) are given by
\[
    \mathcal{D}_m^+  = \big\{ \ket{i}\bra{i} \mid i = 1, \dots, \frac{m}{2}\big\},
\]
and whose negative examples (i.e., labeled $-1$) are given by
\[
    \mathcal{D}_m^-  = \big\{ \ket{i}\bra{i} \mid i = \frac{m}{2} + 1, \dots, m\big\}.
\]
To classify this set of examples we take the classifier $c_{\mathcal{O}, 0} \in \mathcal{C}^{(\eta)}_{\mathrm{qlin}}$ whose observable is given by
\[
    \mathcal{O} = \frac{\eta}{\sqrt{m}}\left(\Big(\sum_{i = 1}^{m/2}\ket{i}\bra{i}\Big) +  \Big(\sum_{j=\frac{m}{2} + 1}^m\ket{j}\bra{j} \Big)\right).
\]
We remark that $c_{\mathcal{O}, 0}$ can indeed classify the set of examples $\mathcal{D}_r$ with margin $\eta/\sqrt{m}$.

Now suppose $c_{\mathcal{O}', b'}  \in \mathcal{C}^{\eta'}_{\text{qlin}}$ with $\eta' < \eta$ can classify $\mathcal{D}_m$ with margin $\gamma'$, that is
\begin{align}
        \mathrm{Tr}\big[\mathcal{O}'\ket{i}\bra{i}\big] \begin{cases}\geq b' + \gamma' &\text{ if }i=1,\dots, \frac{m}{2}, \\ \leq b' - \gamma' &\text{ if }i=\frac{m}{2}+1, \dots, m. \end{cases}
\label{eq:margin}
\end{align}
Define $\rho_+ = \sum_{i = 1}^{m/2}\ket{i}\bra{i}$ and $\rho_- = \sum_{i = \frac{m}{2} + 1}^{m}\ket{i}\bra{i}$ and note that Equation~\eqref{eq:margin} implies that
\[
    \mathrm{Tr}\big[\mathcal{O}'\rho_+\big] \geq \frac{m}{2}b' + \frac{m}{2}\gamma'
\]
and that
\[
    \mathrm{Tr}\big[\mathcal{O}'\rho_-\big] \leq \frac{m}{2}b' - \frac{m}{2}\gamma'
\]
By combining these two inequalities we find that
\begin{align}
     \mathrm{Tr}\big[\mathcal{O}'(\rho_+ - \rho_-)\big] \geq \frac{m}{2}b' - \frac{m}{2}b' + \frac{m}{2}\gamma' + \frac{m}{2}\gamma' = m\gamma'.
     \label{eq:1}
\end{align}
Finally, by the Cauchy–Schwarz inequality we find that
\begin{align}
    \mathrm{Tr}\big[\mathcal{O}'(\rho_+ - \rho_-)\big] \leq \underbrace{||\mathcal{O}'||_F}_{< \eta} \cdot \underbrace{||\rho_+ - \rho_-||_F}_{=\sqrt{m}} < \eta \sqrt{m}.
    \label{eq:cs}
\end{align}
Combining Equation~\eqref{eq:1} and~\eqref{eq:cs} we find that
\[
m\gamma' \leq \mathrm{Tr}\big[\mathcal{O}'(\rho_+ - \rho_-)\big] < \eta \sqrt{m}
\]
from which we can conclude that $\gamma' < \eta/\sqrt{m}$.

\end{proof}
\end{document}